\documentclass[12pt,a4paper]{article}
\usepackage{mathtools}
\usepackage{amsmath}
\usepackage{graphicx,psfrag,epsf}
\usepackage{enumerate}
\usepackage{natbib}
\usepackage{url} 
\usepackage{multicol}
\usepackage{pdflscape}
\usepackage{afterpage}
\usepackage{enumitem}
\usepackage{fmtcount}
\usepackage{relsize}

\usepackage{epstopdf}
\usepackage{subcaption}
\usepackage{siunitx}
\usepackage{titling}
\usepackage{xpatch}

\xpatchcmd{\section}{%
  \normalfont\scshape\centering}{%
  \normalfont\scshape}{\typeout{Success}}{\typeout{Failure}}%

\usepackage{txfonts}
\usepackage{cite}
\usepackage{listings}
\usepackage{booktabs}
\usepackage{amssymb, amsmath}
\usepackage{comment}
\usepackage{changebar}
\usepackage{xcolor}
\usepackage{hyperref}
\usepackage{cases}
\usepackage{courier}
\usepackage{accents}
\usepackage{enumitem}
\usepackage{multirow}
\usepackage{tabularx}
\usepackage{array}
\usepackage{tikz}
\usepackage{bbold}
\usepackage{algpseudocode}
\usepackage[ruled,vlined]{algorithm2e}
\usepackage{setspace}

\usetikzlibrary{fit}
\tikzset{%
  highlight/.style={rectangle,rounded corners,draw,
    fill opacity=0.5,thick,inner sep=0pt}
}

\def\code#1{\texttt{#1}}
\newcommand\numberthis{\addtocounter{equation}{1}\tag{\theequation}}

\DeclarePairedDelimiter{\floor}{\Big\lfloor}{\Big\rfloor}
\DeclareMathOperator*{\argmax}{\arg\!\max}

\newtheorem{Thm}{\bf Theorem}
\newtheorem{Assume}{\bf Assumption}
\newtheorem{Proof}{\bf Proof of Theorem}

\newtheorem{Cor}{\bf Corollary}

\newtheorem{Lem}{Lemma}

\def\b{\boldsymbol}

\begin{document}
\sloppy

\def\spacingset#1{\renewcommand{\baselinestretch}%
{#1}\small\normalsize} \spacingset{0.8}

\title{\bf Collective anomaly detection in High-dimensional VAR Models}

\author{Hyeyoung Maeng, Idris Eckley and Paul Fearnhead\hspace{.2cm} \\
    Lancaster University, United Kingdom}
  \maketitle

\bigskip
\begin{abstract}
There is increasing interest in detecting collective anomalies: potentially short periods of time where the features of data change before reverting back to normal behaviour. We propose a new method for detecting a collective anomaly in VAR models.  Our focus is on situations where the change in the VAR coefficient matrix at an anomaly is sparse, i.e.\ a small number of entries of the VAR coefficient matrix change. To tackle this problem, we propose a test statistic for a local segment that is built on the lasso estimator of the change in model parameters. This enables us to detect a sparse change more efficiently and our lasso-based approach becomes especially advantageous when the anomalous interval is short. We show that the new procedure controls Type 1 error and has asymptotic power tending to one. The practicality of our approach is demonstrated through simulations and two data examples, involving New York taxi trip data and EEG data.
\end{abstract}

\noindent%
{\it Keywords:} Collective anomaly; high-dimensional time series; lasso; sparse changes; epidemic change  
\vfill

\newpage
\spacingset{1.35}

\section{Introduction} \label{sec1}
There is a growing need for modelling and analysis of high-dimensional time series, as such series have become increasingly common in many application areas. 
Applications include 
forecasting using a large panel of time series in economics \citep{de2008forecasting, banbura2010large},
estimating causal relationships among genes and constructing gene regulatory networks \citep{shojaie2010discovering},
identifying the monetary transmission mechanism from macroeconomic time series \citep{bernanke2005measuring},
discovering causal interactions in Neuroimaging \citep{smith2012future, seth2015granger}, 
analysing housing markets for home-price estimation and forecasting \citep{rapach2007forecasting, calomiris2008foreclosure, stock2008evolution} and 
analysing the network structure of volatility interconnections in the Standard \& Poor’s 100 data \citep{barigozzi2017network}.

The majority of existing methods are built on the assumption of stationary and stable time series. If there is either a structural change or a period of anomalous behaviour in a time series, detecting the location of the change/anomaly is not only an important task in itself, but also useful for a follow-up analysis after detection. 
It is indeed a problem of significant interest in many applications.
For example, \citet{chen1997testing} detects multiple change-points in variance of weekly stock prices and \citet{cribben2017estimating} study a network change-point detection problem for resting state functional magnetic resonance imaging  data.
\citet{ombao2005slex} propose a way of segmenting multivariate nonstationary time series and analyse time-varying electroencephalogram data that is recorded during an epileptic seizure.
Other examples include detecting changes that have occurred in a sparse subset of time series \citep{cho2015multiple, cho2016change, wang2018high}, 
covariance change-point detection for multivariate or high-dimensional time series \citep{inclan1994use, aue2009break, wang2017optimal} and
detecting change-points under the factor model framework \citep{breitung2011testing, chen2014detecting, baltagi2017identification, barigozzi2018simultaneous}.

One of the most popular models for high-dimensional time series is the vector autoregressive (VAR) model  \citep{sims1980macroeconomics, lutkepohl2005new}, due to its ability to capture complex temporal and cross-sectional relationships.
However, the estimation of the coefficient matrix becomes challenging as the number of parameters increases quadratically with the number of  time series.
To overcome this, structured sparsity of the VAR coefficients is often assumed as this assumption dramatically reduces the number of model parameters. 
For example, \citet{song2011large} use lasso type, that is $\ell_1$, penalties to encourage sparsity in the estimates of the VAR coefficients.
\citet{davis2016sparse} propose a two-stage approach to fit sparse VAR models and provide a numerical evidence that a log-likelihood based loss function improves the forecasting performance compared to a least squares based one as the former includes information on the error covariance matrix. 
\citet{basu2015regularized} investigate the theoretical properties of $\ell_1$-penalised  estimators for a Gaussian VAR model and show consistency results, while \citet{lin2017regularized} generalise the results by considering a general norm instead of being restricted to the $\ell_1$-norm for the penalty.
Recently, more complex structures have been studied in the literature: \citet{basu2019low} study the low-rank and structured sparse VAR model and \citet{nicholson2020high} impose a hierarchical structure on VAR coefficient matrices according to the lag order, thus addressing both the dimensionality and the lag selection issues at the same time. 

Despite the large body of literature on VAR models, detecting a structural break has rarely been studied. 
\citet{kirch2015detection} consider two scenarios, detecting at-most-one-change and epidemic change in model parameters of multivariate time series which is not restricted to VAR models. 
\citet{safikhani2020joint} consider the multiple change-point setting for the VAR coefficient matrix under a high-dimensional regime and propose a three-stage procedure that returns consistent estimators of both change-points and parameters. 
\citet{wang2019localizing} also study the same setting (i.e.\ when the model parameters have a form of piecewise constant over time) and  use a dynamic programming approach for localising change-points and improving the corresponding error rates. 
\citet{bai2020multiple} study the multiple change-point setting but assume the low-rank plus sparse structure on the VAR coefficient matrices and consider the case where only the sparse structure changes over time, while the low-rank parts remain constant.
We will explain how our proposal is different from these existing works later in this section.

In contrast to these earlier works, we focus on settings where we have plenty of information about the current or normal behaviour of our time-series, and wish to detect periods of different or abnormal behaviour. This can arise when detecting collective anomalies \citep{fisch2018linear,tveten2020scalable} or epidemic changepoints \citep{yao1993tests} -- where we have a, potentially short, period of time where the behaviour of our model changes before it reverts back to current behaviour. This also arises with sequential change detection \citep{lai1995sequential}, when we observe data in real-time and wish to detect any change away from the current behaviour as quickly as possible. For ease of presentation we focus primarily on detecting collective anomalies/epidemic changes, and use the terminology collective anomaly from now on. We show how our method can be extended to the online framework in Section \ref{sec5}. The key feature of these problems is that we have substantially more information about the current or normal behaviour than about the anomaly. This suggests that we should potentially use different procedures to estimate the parameters of the VAR model for the normal behaviour than for the anomaly. We do this through making an assumption that it is the change in VAR parameters that is sparse.


We focus on improving the detection power when the difference between the coefficient matrices at anomaly point is sparse (i.e. a small number of entries of the VAR coefficient matrix change). 
To tackle this problem, we propose a test statistic for a local segment which is built on the lasso estimator of the change in model parameters. 
This enables us to detect a sparse change more efficiently, as the sparsity of change is considered in establishing the test statistic.
Moreover, our lasso-based approach become more advantageous over, say, the standard likelihood-ratio test statistic for shorter anomalous intervals: as for shorter intervals we have fewer observations to estimate the new VAR coefficient matrix, and it becomes more like a high-dimensional problem where the number of observations is similar to or less than the number of parameters to estimate. 
In Section \ref{sec4}, our approach is compared with a method that is built on estimating the change in VAR matrix using ordinary least squares estimator, and the results show that our method outperforms it when detecting sparse change.
As we consider the setting where a relatively longer region has a normal behaviour than the anomalous behaviour, it is reasonable to assume that the underlying VAR coefficient matrix is estimated well enough. Thus, we first develop our method when the normal behaviour is assumed to be known and extend it to the case where an appropriate estimator for the VAR coefficient is used instead. Our theory in Section \ref{sec3} shows the validity of this approach providing that the estimator for the VAR coefficient is close enough to the true one. 
Although our main focus is on single anomaly detection, we show that the new method can be extended for detecting multiple anomalies in Section \ref{sec2.1}.

Among those relevant works already introduced earlier in this section, the work of \citet{safikhani2020joint} and \citet{bai2020multiple} are most closely related to our work, in that they also control the change in VAR parameters with a lasso penalty in their objective functions, however their approaches are different from our method in several aspects. 
To obtain the initial estimate of change-points before screening, \citet{safikhani2020joint} use a fused lasso penalty on a full model considering all time points being a candidate for change-point. Thus their objective function controls the sparsity of VAR parameters and the sparsity of its difference at the same time. 
\citet{bai2020multiple} follows a similar procedure to \citet{safikhani2020joint} under the multiple change-point framework. They use a block fused lasso penalty by assuming that the model parameters in a block is fixed, while our objective function controls only the sparsity of change in building a test statistic and search many segments to find an anomalous interval. 
Also, \citet{safikhani2020joint} and \citet{bai2020multiple} assume that the $l_2$-norm of a change in VAR parameter is bounded away from zero, whereas our assumption on the $l_2$-norm of a change is related to the sparsity of change which is in line with the assumptions used in \citet{wang2019localizing}.
Although those change-point detection methods are not exactly designed for the anomaly setting we consider in this paper, we compare our performance with theirs and present results in the supplementary material. 
Our method works better especially when the underlying VAR coefficient matrix is dense but the change is sparse, and surprisingly even in the case where the VAR coefficient matrix has a low rank plus sparse structure and only a sparse component changes. Full details can be found in the supplementary material.


The remainder of the article is organised as follows. Section \ref{sec2} gives a full description of our procedure and the relevant theoretical results are presented in Section \ref{sec3}. The supporting simulation studies are described in Section \ref{sec4}. Our methodology is illustrated through two datasets in Section \ref{sec5} and we end with additional discussion in Section \ref{sec6}. The proofs of our main theoretical results are in the supplementary material.


\section{Methodology} \label{sec2}
\subsection{Problem setting} \label{sec2.1}
We consider a zero-mean stationary ${p}$-dimensional multivariate time series $\b{x}_t = (x_{1t}, \ldots, x_{pt})'$ generated by a VAR(1) model:
\begin{equation} \label{model}
\b{x}_t = \b{A}_t \b{x}_{t-1} + {\b{\varepsilon}}_t, \quad \b{\varepsilon}_t  \stackrel{\text{i.i.d.}}{\sim} N(\b{0}, \Sigma_{\varepsilon}), \quad t=1, \ldots, T, 
\end{equation}
where $\{\b{A}_{t}\}_{t=1}^{T}$ is a $p \times p$ matrix and $\Sigma_{\varepsilon}$ is a positive definite matrix. 
We assume that the high-dimensional VAR model shows an anomalous behaviour at $t \in [\eta_1, \eta_2]$ such that
\begin{equation}
0=\eta_0 < \eta_1 < \eta_2 < \eta_3=T,
\end{equation}
which gives the sets 
\begin{equation} \label{sets}
\b{\mathfrak{x}}_1 = \{\b{x}_1, \ldots, \b{x}_{\eta_1-1}\}, \quad \b{\mathfrak{x}}_2 = \{\b{x}_{\eta_1}, \ldots, \b{x}_{\eta_2}\}, \quad \b{\mathfrak{x}}_3 = \{\b{x}_{\eta_2+1}, \ldots, \b{x}_{T}\}
\end{equation}
and the sequence of $\{\b{A}_t\}_{t=1}^T$ forms piecewise-constant coefficient matrices as follows,
\begin{align*} 
\b{A}^{(1)} = \b{A}_{1}  = \cdots = \b{A}_{\eta_1-1}, \quad \b{A}^{(2)} = \b{A}_{\eta_1} =\cdots = \b{A}_{\eta_2}, \quad \b{A}^{(1)} = \b{A}_{\eta_2+1} = \cdots \b{A}_{T},
\end{align*}
where $\b{A}^{(1)} \neq \b{A}^{(2)}$ and $\b{A}^{(1)}, \b{A}^{(2)} \in \mathbb{R}^{p \times p}$.

The model in equation \eqref{model} can be represented as the following linear regression,
\begin{equation} \label{model1}
\renewcommand*{\arraystretch}{0.8}
\begin{pmatrix}
\b{x}_1^\prime \\
\b{x}_2^\prime \\
\vdots  \\
\b{x}_T^\prime 
\end{pmatrix}_{T \times p} = 
\begin{pmatrix}
\b{x}_0^\prime & 0 \\
\vdots  & \vdots   \\
\b{x}_{\eta_1-2}^\prime  & 0  \\
\b{x}_{\eta_1-1}^\prime  & \b{x}_{\eta_1-1}^\prime   \\
\vdots  & \vdots   \\
\b{x}_{\eta_2-1}^\prime & \b{x}_{\eta_2-1}^\prime \\
\b{x}_{\eta_2}^\prime & 0 \\
\vdots  & \vdots   \\
\b{x}_{T-1}^\prime  & 0 
\end{pmatrix}_{T \times 2p} 
\begin{pmatrix}
{\b{\theta}^{(1)}}^{\prime} \\
{\b{\theta}^{(2)}}^{\prime} 
\end{pmatrix}_{2p \times p} + 
\begin{pmatrix}
\varepsilon_1^\prime \\
\varepsilon_2^\prime \\
\vdots  \\
\varepsilon_T^\prime 
\end{pmatrix}_{T \times p},
\end{equation}
where 
\begin{align*} 
\b{\theta}^{(1)} = \b{A}^{(1)}, \; \b{\theta}^{(2)} = \b{A}^{(2)}-\b{A}^{(1)}.
\end{align*}
The model, as written in equation \eqref{model1}, is a linear regression of the form $\mathcal{Y} = \mathcal{X} \Theta + {E}.$ 
As such, it can be represented as 
\begin{equation} \label{model4}
\b{Y}_{Tp \times 1}=\b{X}_{Tp \times 2p^2}\b{\Theta}_{2p^2 \times 1} +\b{E}_{Tp \times 1},
\end{equation}
where $\b{X} = \mathit{I}_{p} \otimes \mathcal{X}$ and $\otimes$ is the tensor product of two matrices.

Now our interest is in estimating the collective anomaly $[\eta_1, \eta_2]$. Our motivation is for scenarios where there is substantial information about the normal or pre-change behaviour of the data. Thus, for ease of presentation, we will first assume that $\b{\theta}^{(1)}$ in \eqref{model1} is known. In practice we will use an estimate of $\b{\theta}^{(1)}$, and our theory shows that our approach has good asymptotic properties if we plug-in a suitably accurate estimate of $\b{\theta}^{(1)}$ in the following procedure. We assume that the change $\b{\theta}^{(2)}$ is sparse in that it has small number of nonzero entries which will be formulated in a later section. 
Assuming the base coefficient matrix $\b{A}^{(1)}$ is known, we can rewrite the model as 
\begin{equation} \label{model_a1}
\renewcommand*{\arraystretch}{0.8}
\begin{pmatrix}
\b{x}_1^\prime \\
\b{x}_2^\prime \\
\vdots  \\
\b{x}_T^\prime 
\end{pmatrix}_{T \times p} -
\begin{pmatrix}
\b{x}_0^\prime{\b{\theta}^{(1)}}^{\prime}    \\
\vdots     \\
\b{x}_{\eta_1-2}^\prime{\b{\theta}^{(1)}}^{\prime}     \\
\b{x}_{\eta_1-1}^\prime{\b{\theta}^{(1)}}^{\prime}   \\
\vdots  \\
\b{x}_{\eta_2-1}^\prime{\b{\theta}^{(1)}}^{\prime}  \\
\b{x}_{\eta_2}^\prime{\b{\theta}^{(1)}}^{\prime}    \\
\vdots     \\
\b{x}_{T-1}^\prime{\b{\theta}^{(1)}}^{\prime} 
\end{pmatrix}_{T \times p}
= 
\begin{pmatrix}
 0 \\
\vdots   \\
 0  \\
\b{x}_{\eta_1-1}^\prime   \\
 \vdots   \\
 \b{x}_{\eta_2-1}^\prime \\
  0 \\
\vdots   \\
 0  
\end{pmatrix}_{T \times p} 
\begin{pmatrix}
{\b{\theta}^{(2)}}^{\prime} 
\end{pmatrix}_{p \times p}  + 
\begin{pmatrix}
\varepsilon_1^\prime \\
\varepsilon_2^\prime \\
\vdots  \\
\varepsilon_T^\prime 
\end{pmatrix}_{T \times p},
\end{equation}
which can be represented as $\mathcal{Y} - \mathcal{X}^{(1)}{\b{\theta}^{(1)}}^{\prime} = \mathcal{X}^{(2)}{\b{\theta}^{(2)}}^{\prime}  + {E}.$
With slight abuse of notation as we are using different definitions of $\b{Y}$, $\b{X}$ and $\b{\Theta}$, this can be re-written as 
\begin{equation} \label{model_a2}
\b{Y} _{Tp \times 1}=\b{X} _{Tp \times p^2}\b{\Theta} _{p^2 \times 1} +\b{E} _{Tp \times 1},
\end{equation}
where $\b{X} = \mathit{I}_{p} \otimes \mathcal{X}^{(2)}$. 




\subsection{Lasso-based approach} \label{sec2.2}

To detect a collective anomaly we derive a test for whether data in an interval of time is anomalous, and then apply this test to data from a set of suitably chosen intervals, $\mathbb{J}_{T, p} (L)$. To help with the presentation of theory in Section \ref{sec3}, we parameterise this set by the length, $L$, of the smallest interval it contains. 
For any interval $J \in \mathbb{J}_{T, p} (L)$, by extracting the corresponding rows from each matrix in \eqref{model_a1}, the linear regression form can be rewritten as: $\mathcal{Y}_{J} - \mathcal{X}^{(1)}_{J}{\b{\theta}^{(1)}}^{\prime} = \mathcal{X}^{(2)}_{J}{\b{\theta}^{(2)}}^{\prime}  + {E}_{J},$
that can be vectorised in a form of 
\begin{equation}
    \b{Y}_{J}=\b{X}_{J}\b{\Theta} +\b{E}_{J},
\end{equation}
as in \eqref{model_a2}.

One of the standard ways to detect change or epidemic changes in regression models is to use a likelihood ratio test \citep{kim1989likelihood, siegmund1995using, yau2016inference, baranowski2019narrowest, dette2020likelihood}, and these methods can be applied in the VAR setting. To detect a collective anomaly in a set of intervals, our procedure involves calculating the likelihood ratio statistic for each interval $J \in \mathbb{J}_{T, p} (L)$ as
\begin{equation} \label{lrt}
-2\bigg\{\sum_{s\in J} l_s\big(\b{\Theta}=0, \Sigma_{\varepsilon} \big) - \sum_{s\in J} l_s\big(\b{\hat{\b{\Theta}}}, \Sigma_{\varepsilon} \big)\bigg\}, 
\end{equation}
where $\b{\hat{\b{\Theta}}}$ is the maximum likelihood estimator and the likelihood function has the form of 
\begin{equation*}
\sum_{s\in J} l_s \big(\b{\b{\Theta}}, \Sigma_{\varepsilon} \big) = - \frac{1}{2}\bigg\{ |J|p \log(2 \pi) + |J| \log|\Sigma_{\varepsilon}| + 
(\b{Y}_{J}-\b{X}_{J}{\b{\Theta}})^\top (\Sigma_{\varepsilon}^{-1} \otimes \mathit{I}) (\b{Y}_{J}-\b{X}_{J}{\b{\Theta}}) \bigg\}.
\end{equation*}
As we consider only ${\b{\Theta}}$ varying, the first two terms are constant and will cancel in the test statistic. 


It is common to assume $\Sigma_{\varepsilon}$ is the identity matrix, in which case the maximum likelihood estimator of $\b{\Theta}$ is the ordinary least squares (OLS) estimator. Alternatively we can estimate the variance from the residuals obtained when estimating the parameters of the VAR model on training data.
For ease of presentation, we will assume $\Sigma_{\varepsilon}$ is an identity matrix from now on, but our theoretical results are still valid if this assumption is not correct. Furthermore, the theory can be extended to situations where we assume either $\Sigma_{\varepsilon}$ is any positive identity matrix or an estimate of $\Sigma_{\varepsilon}$ is used. 
We now give details of the likelihood ratio statistic and our suggested improvement based on penalised estimation of the change in the VAR coefficients. 

\paragraph{The OLS method}
Before introducing the lasso-based approach, we consider the test statistic based on the least squares estimator which we refer to as the OLS method. The OLS estimator has been popularly used in the change point detection literature e.g. in a linear model setup, CUSUM-type approaches built on the least squares estimator are studied by \citet{horvath2004monitoring}, \citet{aue2006change}, \citet{chen2010modified} and \citet{fremdt2015page}. For any interval $J \in \mathbb{J}_{T, p} (L)$, the test statistic of the OLS method takes the form, 
\begin{align*} 
T(J) &=  \|\b{Y}_{J} \|_2^2 -  \min_{\b{\Theta}} \big\{ \|\b{Y}_{J}-\b{X}_{J} {\b{\Theta}} \|_2^2\big\}\\ \numberthis \label{naive}
&= \|\b{Y}_{J} \|_2^2- \|\b{Y}_{J}-\b{X}_{J}\hat{\b{\Theta}} \|_2^2,
\end{align*}
that is the same as the likelihood ratio statistic in \eqref{lrt} when $\Sigma_{\varepsilon}$ is the identity matrix.
$T(J)$ has a $\chi^2_{p^2}$ distribution under the null, $\b{\Theta}=\b{0}$. 
The classical least squares estimator $\hat{\b{\Theta}}$ in \eqref{naive} is not able to be used when the dimension $p$ is greater than $T$. 
Note that $\hat{\b{\Theta}}$ also depends on $J$ but this is suppressed in the notation  for simplicity.

\paragraph{The Lasso method}
To handle the case when $\b{\Theta}$ is sparse more effectively, we propose a test statistic based on a lasso estimator:
\begin{equation} \label{lasso}
T^{\text{lasso}}(J) = \|\b{Y}_{J} \|_2^2 -  \min_{\b{\Theta}} \big\{ \|\b{Y}_{J}-\b{X}_{J} {\b{\Theta}} \|_2^2 + \lambda \|\b{\Theta}\|_1 \big\}.
\end{equation}
To detect a collective anomaly, we calculate this test statistic for a collection of intervals, $\mathbb{J}_{T, p} (L)$. We detect an anomaly if the maximum value of these test statistics is above a pre-determined threshold. If we detect an anomaly, we estimate its location as the interval in $\mathbb{J}_{T, p} (L)$ with the largest test-statistic value. The detailed procedure is given in Algorithm \ref{algo1}. 



\begin{center}
\begin{algorithm}[h!]
\setstretch{1}
\SetAlgoLined
\textbf{INPUT}: $\b{X}$ matrix in \eqref{model_a2}, $L$, $\lambda^\textsuperscript{thr}$
        \begin{enumerate}   
            \item[] \textbf{Step 1}: Set a collection of intervals $\mathbb{J}_{T, p} (L)$ where $L$ is the minimum length of intervals.
           \item[] \textbf{Step 2}: For any interval $J \in \mathbb{J}_{T, p} (L)$, calculate $T^{\text{lasso}}(J)$ as in \eqref{lasso}.
           \item[] \textbf{Step 3}: Using a pre-specified threshold $\lambda^\textsuperscript{thr}$, pick the candidate set 
\begin{equation*} 
\mathbb{I}^* = \Big\{ J \in \mathbb{J}_{T, p} (L) : T^{\text{lasso}}(J) > \lambda^\textsuperscript{thr} \Big\}.
\end{equation*}
      \end{enumerate}
If $\mathbb{I}^* \neq \emptyset$, reject the null hypothesis (no anomaly exists) and save the estimator of the anomaly interval,
\begin{equation} \label{ihat}
\hat{I} = \argmax_{ J \in \mathbb{J}_{T, p} (L)} T^{\text{lasso}}(J).
\end{equation}

\textbf{OUTPUT}: $\hat{I}$.
 \caption{Single anomaly detection}
 \label{algo1}
\end{algorithm} 
\end{center}

For setting the collection of intervals $\mathbb{J}_{T, p} (L)$ in Step 1, there exist two general methods; randomly generated intervals \citep{fryzlewicz2014wild, baranowski2019narrowest} and deterministic construction of intervals \citep{kovacs2020seeded}. 
In this paper, we use both construction methods and compare their performance in Section \ref{sec4}.

\subsection{Extension to detecting multiple anomalies} \label{sec2.3}
Following the ideas in \citet{fryzlewicz2014wild} and \citet{kovacs2020seeded}, to deal with multiple anomalies, we repeatedly update the candidate set by removing the intervals that overlap with any detected anomalies. The detailed procedure is given in Algorithm \ref{algo2}.

\begin{center}
\begin{algorithm}[ht]
\setstretch{1}
\SetAlgoLined
\textbf{INPUT}: $\b{X}$ matrix in \eqref{model_a2}, $L$, $\lambda^\textsuperscript{thr}$
        \begin{enumerate}   
            \item[] \textbf{Step 1}: Set a collection of intervals $\mathbb{J}_{T, p} (L)$ where $L$ is the minimum length of intervals.
           \item[] \textbf{Step 2}: For any interval $J \in \mathbb{J}_{T, p} (L)$, calculate $T^{\text{lasso}}(J)$ as in \eqref{lasso}.
           \item[] \textbf{Step 3}: Using a pre-specified threshold $\lambda^\textsuperscript{thr}$, pick the candidate set 
\begin{equation*} \label{candidate}
\mathbb{I}^* = \Big\{ J \in \mathbb{J}_{T, p} (L) : T^{\text{lasso}}(J) > \lambda^\textsuperscript{thr} \Big\}.
\end{equation*}
\end{enumerate}
If $\mathbb{I}^* \neq \emptyset$, reject the null hypothesis (no anomaly exist). Set $\mathbb{I}^{(1)} = \mathbb{I}^*$, $j=1$ and proceed the following steps. \\
\While{$\mathbb{I}^{(j)} \neq \emptyset$}{
  \begin{enumerate}   
\item[] \textbf{Step 4}: Save the estimator of the anomaly interval,
\begin{equation*}
\hat{I}_j = \argmax_{ J \in \mathbb{I}^{(j)} } T^{\text{lasso}}(J),
\end{equation*}
and update the candidate set as 
\begin{equation*}
\mathbb{I}^{(j+1)} = \mathbb{I}^{(j)}  \setminus \{J: J \in \mathbb{I}^{(j)}, J \cap \hat{I}_j \neq \emptyset\}
\end{equation*}
\item[] \textbf{Step 5}: Set $j=j+1$.
      \end{enumerate}
 }
\textbf{OUTPUT}: $\hat{I} = \{ \hat{I}_1, \hat{I}_2, \cdots\}$.
 \caption{Multiple anomaly detection}
 \label{algo2}
\end{algorithm} 
\end{center}

\subsection{Extension to VAR(q) model} \label{sec2.4}
The VAR process of order $1$ presented in Section \ref{sec2.1} can simply be extended to VAR(q) as follows,
\begin{equation} \label{model_varq}
\b{x}_t = \b{A}_{t, 1} \b{x}_{t-1} + \cdots + \b{A}_{t, q} \b{x}_{t-q} + {\b{\varepsilon}}_t, \quad \b{\varepsilon}_t  \stackrel{\text{i.i.d.}}{\sim} N(\b{0}, \Sigma_{\varepsilon}), \quad t=1, \ldots, T, 
\end{equation}
where $\{\b{A}_{t, k}\}_{t=1}^{T}$ is a $p \times p$ matrix for all $k=1, \ldots, q$ and $\Sigma_{\varepsilon}$ is assumed to be a positive definite matrix. With a slight abuse of notation, the piecewise-constant coefficient matrices are as follows,
\begin{align*} 
&\b{A}^{(1)} = (\b{A}_{t', 1}, \ldots, \b{A}_{t', q}) \in \mathbb{R}^{p \times pq}, \quad \text{ for any } t'=1, \ldots, {\eta_1-1}, {\eta_2+1}, \ldots, T \\
&\b{A}^{(2)} = (\b{A}_{t', 1}, \ldots, \b{A}_{t', q}) \in \mathbb{R}^{p \times pq}, \quad \text{ for any } t'= {\eta_1}, \ldots, {\eta_2},
\end{align*}
and the model \eqref{model_varq} can be represented as
\begin{equation} \label{model1_varq}
\renewcommand*{\arraystretch}{0.6}
\begin{pmatrix}
\b{x}_q^\prime \\
\b{x}_{q+1}^\prime \\
\vdots  \\
\b{x}_{T}^\prime 
\end{pmatrix}_{(T-q+1) \times p} = 
\begin{pmatrix}
\b{x}_{q-1}^\prime & \cdots & \b{x}_0^\prime & 0 & \cdots & 0\\
\vdots & & \vdots  & \vdots & & \vdots \\
\b{x}_{\eta_1+q-3}^\prime & \cdots & \b{x}_{\eta_1-2}^\prime  & 0  & \cdots & 0\\
\b{x}_{\eta_1+q-2}^\prime & \cdots & \b{x}_{\eta_1-1}^\prime  & \b{x}_{\eta_1+q-2}^\prime & \cdots & \b{x}_{\eta_1-1}^\prime   \\
\vdots & & \vdots  & \vdots & & \vdots \\
\b{x}_{\eta_2+q-2}^\prime & \cdots & \b{x}_{\eta_2-1}^\prime & \b{x}_{\eta_2+q-2}^\prime & \cdots & \b{x}_{\eta_2-1}^\prime\\
\b{x}_{\eta_2+q-1}^\prime & \cdots & \b{x}_{\eta_2}^\prime & 0 & \cdots & 0\\
\vdots & & \vdots  & \vdots & & \vdots \\
\b{x}_{T-1}^\prime & \cdots & \b{x}_{T-q}^\prime & 0 & \cdots & 0\\
\end{pmatrix}_{(T-q+1) \times 2pq} 
\begin{pmatrix}
{\b{\theta}^{(1)}}^{\prime} \\
{\b{\theta}^{(2)}}^{\prime} 
\end{pmatrix}_{2pq \times p} + 
\begin{pmatrix}
\varepsilon_q^\prime \\
\varepsilon_{q+1}^\prime \\
\vdots  \\
\varepsilon_T^\prime 
\end{pmatrix}_{(T-q+1) \times p},
\end{equation}
where $\b{\theta}^{(1)} = \b{A}^{(1)}, \b{\theta}^{(2)} = \b{A}^{(2)}-\b{A}^{(1)}$.
With the larger dimension of the parameters, the same argument for the VAR process of order $q$ can be achieved by following the logic from \eqref{model4}.

\section{Theoretical results} \label{sec3}
In this section, we explore the asymptotic behaviour of the proposed method. 
We show that our method controls the familywise error under the null (i.e. when there exist no anomaly) with an appropriate threshold and give conditions under which the asymptotic power of the method tends to 1. 
These results are based upon the following assumptions. 

\begin{Assume} \label{asmpt1}
For each $j=1, 2$, let $\Gamma_j(\ell)$ be the population version of the lag-$\ell$ covariance matrix of $\b{\mathfrak{x}}_j$ where $\b{\mathfrak{x}}_j$ is as in \eqref{sets}.
For $\kappa \in [-\pi, \pi]$, there exist the spectral density matrices,
\begin{equation*}
f_j(\kappa) = \frac{1}{2\pi}\sum_{l \in \mathbb{Z}} \Gamma_j(l) \exp^{-\sqrt{-1}\kappa l}.
\end{equation*}
In addition, 
\begin{equation*}
\max_j \mathcal{M}(f_j) = \max_j \Big\{ \text{ess}\sup_{\kappa\in [-\pi, \pi]} \Lambda_{\max}(f_j(\kappa)) \Big\} < +\infty,
\end{equation*}
and
\begin{equation*}
\min_j \b{\mathfrak{m}}(f_j) = \min_j \Big\{ \text{ess}\inf_{\kappa\in [-\pi, \pi]} \Lambda_{\min}(f_j(\kappa)) \Big\} > 0,
\end{equation*}
where $\Lambda_{\max}(A)$ and $\Lambda_{\min}$ are the largest and the smallest eigenvalues of the symmetric matrix $A$, respectively.
\end{Assume}
This first condition is needed to control the stability properties of the VAR models.
This is a spectral density condition that is not only valid for VAR model but also holds for a large class of general linear process. \citet{basu2015regularized} use the same assumption but for a stable VAR setting without considering anomalies, while we extend it to the single collective anomaly setting by assuming a spectral density function for each common and anomalous segments separately.


In order to bound the power of our method we need conditions on the size and length of any anomaly and the set of intervals we use -- essentially we will need at least one interval of sufficient length to be contained within the anomaly. To this end we introduce the following:
\begin{Assume} \label{asmpt2}
There exist at least one interval $J \in \mathbb{J}_{T, p} (L)$ such that $J \subseteq [\eta_1, \eta_2]$ and the choice of $L$ for a set of intervals $\mathbb{J}_{T, p} (L)$ satisfies the following condition as $T, p \rightarrow \infty$,
\begin{equation*}
\frac{log(T \lor p)}{L} \rightarrow 0
\end{equation*}
where any interval $J \in \mathbb{J}_{T, p} (L)$ has length at least $L$.
\end{Assume}


\begin{Assume} \label{asmpt3}
The sparsity of change is fixed; $\|\b{\Theta} \|_0=d_0$. 
\end{Assume}

\begin{Assume} \label{asmpt4}
For any $\xi>0$, $L \cdot \|\b{\Theta}\|_2^2  > C_2 \cdot d_0^2 \cdot \log^{1+\xi}{(T \lor p)}$, where $C_2>0$ is a constant.
\end{Assume}
Assumption \ref{asmpt3} gives the condition on the number of nonzero entries of the coefficient matrix, where the sparsity parameter $d_0$ affects the signal-to-noise ratio condition in Assumption \ref{asmpt4}. 
Our Assumption \ref{asmpt4} is similar to the conditions required in other change-point problem in high-dimensional VAR model. For example, \citet{wang2019localizing} study a multiple change point setting and their signal-to-noise ratio assumption becomes equal to ours in the case when single change-point is considered, while \citet{safikhani2020joint} assume $\|\b{\Theta}\|_2$ is bounded away from zero.

Our final assumption is used to extend our results to the case where we estimate $\b{\theta}^{(1)}$.
\begin{Assume} \label{asmpt5}
It holds for the estimator $\hat{\b{\theta}}^{(1)}$ that $\big\|\b{\theta}^{(1)}-\hat{\b{\theta}}^{(1)} \big\|_\infty < C \sqrt{\frac{\log({T \lor p})}{L}}$ with probability approaching $1$ as $T \rightarrow \infty$ and $p \rightarrow \infty$, where $C>0$ is a constant. 
\end{Assume}
Assumption \ref{asmpt5} states the condition on the estimation error bound in $\ell_\infty$-norm. This is in line with the estimation error presented in Proposition 4.1 of \citet{basu2015regularized} and Lemma 15 of \citet{wang2019localizing} in which the sparsity assumption is imposed on VAR coefficient matrices. For instance, when $\b{\theta}^{(1)}$ is assumed to be sparse with the condition $\|\b{\theta}^{(1)}\|_0=k$, then its lasso estimator, $\hat{\b{\theta}}^{(1)}$, satisfies $\big\|\b{\theta}^{(1)}-\hat{\b{\theta}}^{(1)} \big\|_2 \leq c \sqrt{k} \sqrt{\frac{\log ({T \lor p})}{T}}$, where $\hat{\b{\theta}}^{(1)}$ is obtained from a sample of size $T$. When the sparsity $k$ is fixed, the estimation error bound in $\ell_2$-norm implies Assumption \ref{asmpt5}.


We now present our main theoretical results.
The following theorem gives conditions on the lasso penalty to ensure the procedure asymptotically controls the familywise error when there is no anomaly.  
\begin{Thm} \label{theorem1}
Let Assumptions \ref{asmpt1}-\ref{asmpt2} hold. If there exist no anomaly, for a tuning parameter $\lambda = C_3 \sqrt{L (2\log{p} + \log{T})}$ with a constant $C_3$ large enough, we have
\begin{align*}
P\bigg(\max_{ J \in \mathbb{J}_{T, p} (L)} T^{\text{lasso}}(J) \leq \lambda^{\text{thr}} \bigg) 
& \geq P\bigg(\max_{ J \in \mathbb{J}_{T, p} (L)}  T^{\text{lasso}}(J) = 0\bigg) \\ 
& \geq 1-C_4 \exp(-C_5 (2\log{p} + \log{T})),
\end{align*}
where $C_4, C_5 >0$, $\lambda^{\text{thr}}$ is strictly positive and $\lambda$ is a tuning parameter controlling the penalty term in lasso regression in \eqref{lasso}.
\end{Thm}

In Theorem \ref{theorem1}, it is clear that our result applies to any positive threshold $\lambda^{\text{thr}}$. In the proof of Theorem \ref{theorem1}, we show that the familywise error is controlled under an appropriate tuning parameter $\lambda$ 
and the proof can be found in the supplementary material.
We now turn to the asymptotics of the test statistic under the alternative. 


\begin{Thm} \label{theorem2}
Let Assumptions \ref{asmpt1}-\ref{asmpt4} hold. 
If there exist an anomaly, with a tuning parameter $\lambda = C_2 \sqrt{L (2\log{p} + \log{T})}$ for a large enough $C_2$, as $T \rightarrow \infty$,
\begin{align*}
P\bigg(\max_{ J \in \mathbb{J}_{T, p} (L)} T^{\text{lasso}}(J) \leq \lambda^{\text{thr}} \bigg)  \rightarrow 0
\end{align*}
and 
\begin{equation*}
P(\hat{I} \cap [\eta_1, \eta_2] \neq \emptyset) \rightarrow 1,
\end{equation*}
where the threshold $\lambda^{\text{thr}}$ has the order of $\sqrt{L\cdot\log (p \lor T)}$, the estimated anomaly $\hat{I}$ is as in \eqref{ihat} and $\lambda$ is a tuning parameter controlling the penalty term in lasso regression in \eqref{lasso}.
\end{Thm}




Theorem \ref{theorem2} states that the test statistic corresponding to the intervals in the candidate set is greater than the pre-specified threshold if the interval is located within the true anomaly. In other words, it shows that the individual test has asymptotic power one. 
The following theorem shows that our method has larger power to detect a sparse collective anomaly.


\begin{Thm} \label{thm2-1}
Assume that $\b{x}_t$ follows \eqref{model_a1} and let Assumptions \ref{asmpt1}-\ref{asmpt4} hold. 
Let the null hypothesis hold, then for any $\{J: J \in \mathbb{J}_{T, p} (L), J \cap [\eta_1, \eta_2] = \emptyset \}$, the test statistic of the OLS method in \eqref{naive} follows a $\chi^2_{p^2}$ distribution. 
Consequently, we have an asymptotic level $\alpha$ test if the null hypothesis is rejected for $T(J) > \chi^2_{p^2; (1-\alpha)}$, where $\chi^2_{p^2; (1-\alpha)}$ is the $(1-\alpha)$-quantile of chi-square distribution with $p^2$ degrees of freedom.

Under the alternative, for any $J \in \mathbb{J}_{T, p} (L)$ such that $J \subseteq [\eta_1, \eta_2]$, the upper bound on the power of the OLS method is given as 
\begin{equation} \label{upperbound}
\frac{E\big( \|\b{Y}_{J} \|_2^2  - \|\b{Y}_{J}-\b{X}_{J}{\b{\Theta}} \|_2^2\big)}{W_p},
\end{equation}
where $W_p=O_p(p)$.
\end{Thm}
Note that $W_p$ in \eqref{upperbound} is linked to the false positive rate as it is the approximation of $\chi^2_{p^2; (1-\alpha)} - p^2$. See the proof in the supplementary material for further details.

Theorem \ref{thm2-1} shows the asymptotic behaviours of the test statistic of the OLS method under both the null and the alternative hypotheses.  Furthermore, Theorem \ref{thm2-1} implies that the test statistic built on the lasso estimator can detect weaker anomalies than using the OLS estimator when the change is sparse. 
The intuition behind this is that the test statistic of the OLS method in \eqref{naive} can be written as
\begin{equation} \label{naive_alt}
\|\b{Y}_{J} \|_2^2  - \|\b{Y}_{J}-\b{X}_{J}{\b{\Theta}} \|_2^2 + \big\{ \|\b{Y}_{J}-\b{X}_{J}{\b{\Theta}} \|_2^2 - \|\b{Y}_{J}-\b{X}_{J}\hat{\b{\Theta}} \|_2^2 \big\},
\end{equation}
and $E(\|\b{Y} \|_2^2-\|\b{Y}-\b{X}{\b{\Theta}} \|_2^2)$ needs to be at least as large as $O_p(p)$ to have high power.
By comparison, if we denote the lasso estimator of $\b{\Theta}$ by $\hat{\b{\Theta}}$, then the test statistic of the lasso method in \eqref{lasso} can be written as
\begin{align*} \numberthis \label{lasso_alt}
&\|\b{Y}_{J} \|_2^2  -\|\b{Y}_{J}-\b{X}_{J}{\b{\Theta}} \|_2^2 - \lambda\|\b{\Theta}\|_1 + \big\{ \|\b{Y}_{J}-\b{X}_{J}{\b{\Theta}} \|_2^2 + \lambda\|\b{\Theta}\|_1 - \|\b{Y}_{J}-\b{X}_{J}\hat{\b{\Theta}} \|_2^2 - \lambda\|\hat{\b{\Theta}}\|_1 \big\}.
\end{align*}
Noting that the term in $\{\}$s in \eqref{lasso_alt} is positive, the lasso-based test statistic requires that $\|\b{Y} \|_2^2-\|\b{Y}-\b{X}{\b{\Theta}} \|_2^2$ should at least as large as $O_p(\lambda\|\b{\Theta}\|_1)$ and $\lambda = C_2 \sqrt{L(2\log{p} + \log{T})}$. 


The following two corollaries state that the assertions in Theorems \ref{theorem1}-\ref{theorem2} remain true if the $\b{\theta}^{(1)}$ is replaced by an estimator ${\hat{\b{\theta}}^{(1)}}$ that satisfies the condition in Assumption \ref{asmpt5}.

\begin{Cor} \label{cor1}
Theorem \ref{theorem1} holds with a different constant if ${\hat{\b{\theta}}^{(1)}}$ is used in calculating the test statistic instead of the true parameter $\b{\theta}^{(1)}$, where ${\hat{\b{\theta}}^{(1)}}^{\prime}$ is an estimator fulfilling Assumption \ref{asmpt5}.
\end{Cor}

\begin{Cor} \label{cor2}
Theorem \ref{theorem2} holds with a different constant if ${\hat{\b{\theta}}^{(1)}}$ is used in calculating the test statistic instead of the true parameter $\b{\theta}^{(1)}$, where ${\hat{\b{\theta}}^{(1)}}^{\prime}$ is an estimator fulfilling Assumption \ref{asmpt5}.
\end{Cor}

The proofs of Theorems \ref{theorem1}-\ref{thm2-1} and Corollaries \ref{cor1}-\ref{cor2} can be found in the supplementary material.


\section{Simulation study} \label{sec4}
\subsection{Preliminaries}

We compare the performance of our lasso-based approach with the OLS method described in Section \ref{sec2.2}. Whilst there are other methods for detecting changes in a VAR model, such as those of \citet{safikhani2020joint} and \citet{bai2020multiple}, they are not designed for the collective anomaly setting that we consider. For completeness, 
we compare their performances with ours, and the details can be found in the supplementary material. 
Perhaps due to not being designed for the collective anomaly setting, we find these alternative methods perform substantially worse than ours, particularly when the underlying matrix $A^{(1)}$ is dense but the change is sparse. 

In practice, the underlying parameter $A^{(1)}$ is often unknown and needs to be estimated. 
In this case, as the accuracy of our method depends on how accurately we can estimate $A^{(1)}$, considering two extreme cases gives upper and lower bounds on our method: $A^{(1)}$ is known and $A^{(1)}$ is estimated from a relatively small amount of data. 
%
The threshold of each test is selected by choosing the $99\%$ quantile of the test statistics obtained through the 100 simulation runs performed under the null.
For the error variance, we set $\Sigma_{\varepsilon}$  to be the identity matrix. 
In the following sections, we report the results when $\Sigma_{\varepsilon}$ is known. The results for the case when $\Sigma_{\varepsilon}$ is estimated can be found in the supplementary material.


We also look at how the choice of the set of intervals, $\mathbb{J}_{T, p} (L)$, affects performance. We vary both the number of intervals which we denote by $s$, and the way we choose the intervals, randomly or deterministically, with a pre-determined minimum length of interval.
For the deterministic construction of intervals, we use the technique proposed in Definition 1 of \citet{kovacs2020seeded} with the decay parameter $1/a = 1.1, 1.2$.
Regardless of the way of choosing the intervals, we force the minimum length intervals to be greater than $p$ in order to compare our approach with the OLS method.
In the following sections, we present the simulation results for two scenarios: (1) $A^{(1)}$ is dense and (2) $A^{(1)}$ is sparse; where the number of non-zero elements is large in $(1)$ and small in $(2)$.
Note that when $A^{(1)}$ is assumed to be unknown, it is estimated from the null region with ridge or lasso penalty depending on the given sparsity of $A^{(1)}$.

\subsection{Dense $A^{(1)}$} \label{denseA1}
We first consider the case when all entries of $A^{(1)}$ are non-zero.
The coefficient matrix is randomly generated by using the algorithm proposed by \citet{ansley1986note} and implemented in R package \code{gmvarkit} which forces the resulting VAR model to be stationary, where the range of the entries of $A^{(1)}$ is obtained as $[-0.67, 0.58]$.
We set $T=500$ and $p=10$, where only a few (five or ten) entries in the VAR coefficient matrix undergo change in anomalous interval. 
We investigate both cases: (1) $A^{(1)}$ is assumed to be known and (2) $A^{(1)}$ is estimated from the training data with a ridge penalty.
In the latter case, the training data contains the same amount of the test data which we examine for detecting an anomaly. 
Our lasso-based method is implemented by using the tuning parameter $\lambda$ presented in Theorems \ref{theorem1}-\ref{theorem2} with the constant $C=0.15$.
In the following sections, we consider the single anomaly and the multiple anomaly case.


\subsubsection{Single anomaly} \label{sec.s.a}

\begin{figure}[h!]
\begin{center}
\includegraphics[width=12cm, height=5cm]{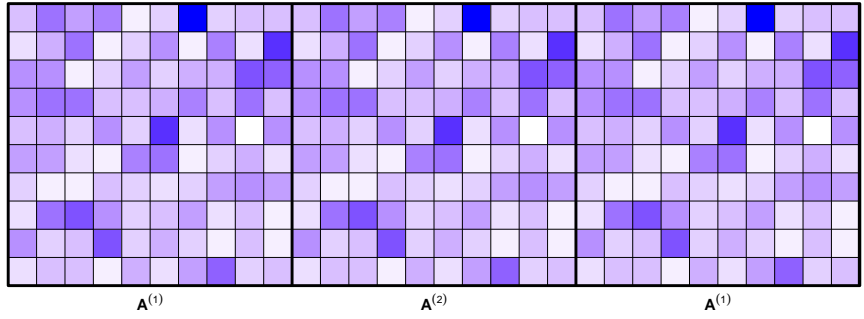}
\end{center}
\caption{The underlying coefficient matrices, $(A^{(1)}, A^{(2)}, A^{(1)})$, for the simulation setting in Section \ref{sec.s.a}, where $A^{(2)}$ corresponds to an anomaly.}
\label{fig:single_a}
\end{figure}

We consider a single anomaly interval located in the middle with three different lengths, where the details can be found in Table \ref{Tab:single_a} and the coefficient matrices are presented in Figure \ref{fig:single_a}.
The non-zero entries in $A^{(2)}-A^{(1)}$ are all equal to $\Delta = 0.35$ and $A^{(2)}$ is made by adding $\Delta$ to the first ten smallest positive entries of $A^{(1)}$. 

\begin{table}[h!] 
\centering
\begin{tabular}{ccccccc} 
  \hline
   & T & p & $[\eta_1, \eta_2]$ & $\eta_2-\eta_1$ & $\Delta$ & $\|\b{\Theta}\|_0$ \\
  \hline
case 1 & 500 & 10 & $[T(5/11), T(6/11)]$ & $45$ & 0.35 & 10 \\
case 2 & 500 & 10 & $[T(7/15), T(8/15)]$ & $33$ & 0.35 & 10 \\
  \hline
\end{tabular} 
\caption{Simulation settings for two cases considered in Section \ref{sec.s.a}, where $\Delta$ is the size of non-zero entries of $\b{\Theta}$ and $\|\b{\Theta}\|_0$ is the number of non-zero elements of $\b{\Theta}$.}
\label{Tab:single_a}
\end{table}

\begin{table}[h!]
\centering
\begin{tabular}{ccccc}
  \hline
 & & & $A^{(1)}$ is known & $A^{(1)}$ is estimated\\
 \hline
 \multirow{6}{*}{case 1} & random  & OLS  &  100  &   82 \\ 
            &  (s = 1029) & Lasso &  100 & \textbf{94}  \\   
            \cline{2-5}  
            & deterministic  & OLS  & 100 &  85 \\ 
            &  (s = 1029) & Lasso &  100 &  \textbf{95}\\   
            \cline{2-5}
            & deterministic  & OLS  &  100  &  85  \\ 
            &  (s = 540) & Lasso &  100&  \textbf{94} \\     
            \hline
\multirow{6}{*}{case 2} & random  & OLS  &  98  & 46   \\ 
            &  (s = 1029) & Lasso &  \textbf{100} &  \textbf{66} \\  
            \cline{2-5}
            & deterministic  & OLS  &  98 &  56\\ 
            &  (s = 1029) & Lasso & \textbf{99} &  \textbf{75}\\        
            \cline{2-5}
            & deterministic  & OLS  &   98 &  54  \\ 
            &  (s = 540) & Lasso &  \textbf{99} &   \textbf{72} \\    
            \hline              
\end{tabular}
\caption{Empirical power ($\%$) from 100 simulation runs for two methods in all cases described in Section \ref{sec.s.a}, where $s$ is the number of intervals examined.}
\label{tab:single_a_count}
\end{table}


\begin{table}[h!]
\centering
\begin{tabular}{ccccc}
  \hline
 & & & $A^{(1)}$ is known & $A^{(1)}$ is estimated\\
 \hline
 \multirow{6}{*}{case 1} & random  & OLS  & 0.45 (0.31)   & 10.50 (16.66)   \\ 
            &  (s = 1029) & Lasso &  0.40 (0.00) & 5.24 (10.34)  \\  
            \cline{2-5}  
            & deterministic  & OLS  & 0.39 (0.25)  &  7.18 (16.31) \\  
            &  (s = 1029) & Lasso & 0.35 (0.16) & 2.63 (9.96) \\      
            \cline{2-5}
            & deterministic  & OLS  &  0.39 (0.33)  &  7.15 (16.32) \\ 
            &  (s = 540) & Lasso &  0.32 (0.22) & 3.06 (10.85) \\ 
            \hline
\multirow{6}{*}{case 2} & random  & OLS  & 3.42 (6.49)   &  28.38 (21.04) \\ 
            &  (s = 1029) & Lasso & 2.20 (0.91) & 19.08 (20.54) \\ 
            \cline{2-5}
            & deterministic  & OLS  &  1.32 (6.57) &  21.31 (23.20)\\ 
            &  (s = 1029) & Lasso & 0.82 (4.67) & 12.46 (20.57) \\       
            \cline{2-5}
            & deterministic  & OLS  &  1.27 (6.58)  & 22.17 (23.34) \\ 
            &  (s = 540) & Lasso & 0.76 (4.68) & 13.76 (21.34) \\              
            \hline    
\end{tabular}
\caption{The mean (standard deviation) of Hausdorff distance from 100 simulation runs for two methods in all cases described in Section \ref{sec.s.a}, where $s$ is the number of intervals examined.}
\label{tab:single_a_meansd}
\end{table}

As shown in Table \ref{tab:single_a_count}, the lasso-based  method tends to detect an anomaly more often than the OLS-based approach in all cases regardless of the way of choosing intervals to investigate and whether $A^{(1)}$ is known or unknown. As expected,
compared to the results when the true $A^{(1)}$ is known, both OLS and lasso methods perform less well when $\hat{A}^{(1)}$ is used.
Comparing the randomly and the deterministically chosen segments with the size ($s$) equal to $1029$, for both the OLS and the lasso methods, the deterministic way tends to give a slightly lower power when $A^{(1)}$ is known but gives a similar or a slightly larger power when $A^{(1)}$ is estimated. 
Note, when $A^{(1)}$ is estimated, the deterministically chosen intervals with smaller sample size ($s=540$) shows a similar or a larger power than those chosen randomly with a sample size ($s=1029$) for both methods, and the difference becomes larger as the length of anomalous interval becomes shorter (from case 1 to case 2 as presented in Table \ref{Tab:single_a}). 
Table \ref{tab:single_a_meansd} shows that the lasso method also outperforms in terms of distance between the estimated and the true anomaly and its variance.





\subsubsection{Two collective anomalies} \label{ta}

\begin{figure}[!ht]
\begin{center}
\includegraphics[width=16cm, height=4.5cm]{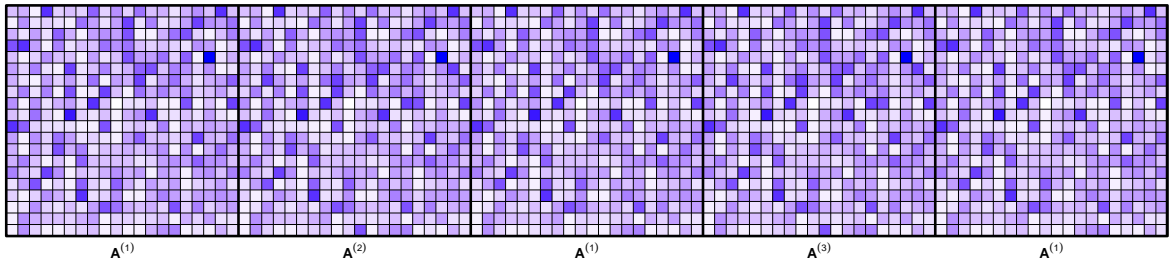}
\end{center}
\caption{The underlying coefficient matrices, $(A^{(1)}, A^{(2)}, A^{(1)}, A^{(3)}, A^{(1)})$, for the simulation setting in Section \ref{ta}, where $A^{(2)}$ and $A^{(3)}$ correspond to the anomalies.}
\label{fig:two_a}
\end{figure}

We now consider two collective anomalies, $[\eta_1, \eta_2]$ and $[\eta_3, \eta_4]$, where the corresponding coefficient matrix is $A^{(2)}$ and $A^{(3)}$, respectively.
$A^{(2)}$ and $A^{(3)}$ are obtained by adding $\Delta_1$ and $\Delta_2$, respectively to the first five smallest positive entries of $A^{(1)}$ and the true coefficient matrices are presented in Figure \ref{fig:two_a}.
Two different cases are considered and the details are provided in Table \ref{table_ta}. 

\begin{table}[!ht]
\centering
\begin{tabular}{cccccccccc}
  \hline
  & T & p  & $[\eta_1, \eta_2]$ & $[\eta_3, \eta_4]$ &  $\eta_2-\eta_1$ & $\eta_4-\eta_3$ &  $\Delta_1$ &  $\Delta_2$ & $\|\b{\Theta}\|_0$ \\
  \hline
case 1 & 500 & 10 & $[133, 166]$ & $[333, 366]$ & $33$ & $33$ & $0.6$ & $0.6$ & $5$ \\
case 2 & 500 & 10 & $[33, 66]$ & $[433, 466]$& $33$ & $33$ & $0.5$ & $0.5$ & $5$ \\
  \hline
\end{tabular}
\caption{Simulation settings for two cases considered in Section \ref{ta}, where $\Delta_1=|A^{(2)}-A^{(1)}|$, $\Delta_2=|A^{(3)}-A^{(1)}|$ and $\|\b{\Theta}\|_0$ is the number of non-zero elements of $\b{\Theta}$.}
\label{table_ta}
\end{table}

\begin{table}[h!]
\centering
\begin{tabular}{cccccccccccc}
  \hline
 & & & \multicolumn{4}{c}{$A^{(1)}$ is known} & & \multicolumn{4}{c}{$A^{(1)}$ is estimated} \\
 \cmidrule(lr){4-7} \cmidrule(lr){9-12} 
 & & & 0 & 1 & \textbf{2} & 3  & & 0 & 1 & \textbf{2} & 3 \\
\hline
\multirow{6}{*}{case 1} & random  & OLS  & 0 & 27& \textbf{73} & 0  && 3 & \textbf{82} & 15 & 0 \\ 
            &  (s = 1944) & Lasso & 0 & 24 & \textbf{76} & 0  && 0 & 48 &\textbf{52} & 0 \\ 
            \cline{2-12}  
            & deterministic  & OLS  & 0 & 24& \textbf{76} & 0 && 1 & \textbf{70} & 29 & 0 \\ 
            &  (s = 1944) & Lasso & 0 & 12 & \textbf{86}  & 2 && 0 & 35  & \textbf{65} & 0 \\ 
            \cline{2-12}
            & deterministic  & OLS  & 0 & 26 & \textbf{74} & 0  && 2 & \textbf{72}  & 26 & 0 \\ 
            &  (s = 1029) & Lasso & 0 & 21 & \textbf{77}  & 2 && 0 & 40 & \textbf{60} & 0 \\ 
            \hline
 \multirow{6}{*}{case 2} & random  & OLS  & 0 & 4 & \textbf{94} & 2 && 36 & \textbf{58} & 6 & 0 \\ 
            &  (s = 1944) & Lasso & 0 & 2  & \textbf{98} & 0 && 7 & \textbf{50} & 43 & 0 \\ 
            \cline{2-12}  
            & deterministic  & OLS  & 0 & 4 & \textbf{95} & 1 && 31 & \textbf{60} & 9 & 0 \\ 
            &  (s = 1944) & Lasso & 0 & 1 & \textbf{96} & 3 && 4 & 43 & \textbf{53} & 0 \\ 
            \cline{2-12}
            & deterministic  & OLS  & 0 & 4 & \textbf{95} & 1 && 34 & \textbf{58}  & 8 & 0 \\ 
            &  (s = 1029) & Lasso & 0 & 1 & \textbf{98} & 1 && 8 & 40 & \textbf{52} & 0 \\ 
            \hline
\end{tabular}
\caption{Distribution of the number of detected anomalies for two methods in all cases described in Section \ref{ta} over 100 simulation runs, where $s$ is the number of intervals examined.}
\label{tab:two_a_count}
\end{table}

\begin{table}[h!]
\centering
\begin{tabular}{ccccc}
  \hline
 & & & $A^{(1)}$ is known & $A^{(1)}$ is estimated\\
 \hline
 \multirow{6}{*}{case 1} & random  & OLS  & 3.01 (2.89)  & 15.90 (12.21)   \\ 
            &  (s =  1944) & Lasso &  2.82 (2.83) & 7.38 (8.56)\\  
            \cline{2-5}  
            & deterministic  & OLS  & 2.46 (2.62)  &  12.45 (13.04)\\  
            &  (s =   1944) & Lasso & 1.97 (2.68) & 4.41 (7.26) \\      
            \cline{2-5}
            & deterministic  & OLS  & 2.59 (2.57) & 13.07 (13.06)  \\ 
            &  (s =  1029) & Lasso &  2.43 (2.66) & 5.11 (7.77)\\ 
            \hline
\multirow{6}{*}{case 2} & random  & OLS  & 3.93 (5.11)   &  12.26 (2.83)\\ 
            &  (s =  1944) & Lasso & 2.50 (1.88) & 8.38 (5.44)  \\ 
            \cline{2-5}
            & deterministic  & OLS  & 1.92 (4.19) &  11.92 (3.76)\\ 
            &  (s = 1944) & Lasso & 1.50 (3.80) & 6.66 (6.22) \\       
            \cline{2-5}
            & deterministic  & OLS  &  1.98 (4.22)  &  12.02 (3.58) \\ 
            &  (s =  1029) & Lasso & 1.44 (3.68) & 6.79 (6.17) \\              
            \hline
\end{tabular}
\caption{The mean (standard deviation) of Hausdorff distance from 100 simulation runs for two methods in all cases described in Section \ref{ta}, where $s$ is the number of intervals examined.}
\label{tab:two_a_meansd}
\end{table}

To detect multiple anomalies, the procedure presented in Algorithm \ref{algo2} is applied.
From Table \ref{tab:two_a_count}, we obtain similar interpretations to those from the single anomaly case in Section \ref{sec.s.a}.
When the true $A^{(1)}$ is known, both the OLS and the lasso methods give better results than when $A^{(1)}$ is assumed to be unknown and estimated. 
For both methods and for both cases ($A^{(1)}$ is known and estimated), the deterministic settings ($s=1029$ and $s=1944$) tend to return better results than the random setting with the sample size of $s=1944$. 
In Table \ref{tab:two_a_meansd}, the lasso method returns smaller mean and standard deviation of the Hausdorff distance, regardless of the way of choosing segments and whether $A^{(1)}$ is known or not.

\subsection{Sparse $A^{(1)}$} \label{sec4.2.1}

\begin{table}[!ht] 
\centering
\begin{tabular}{cccccccc} 
  \hline
          & T & p &  $[\eta_1, \eta_2]$ & $\eta_2-\eta_1$ & $\|A^{(1)}\|_\infty$ & $\|A^{(2)}\|_\infty$  & $\|\b{\Theta}\|_0$ \\
  \hline
case 1 & 500 & 20 & $[T(4/9), T(5/9)]$ & $55$ & $0.6$ & $0.05$ & $19$\\
case 2 & 500 & 20 & $[T(6/13), T(7/13)]$ & $39$ & $0.6$ & $0.05$ & $19$\\
  \hline
\end{tabular} 
\caption{Simulation setting for two cases considered in Section \ref{sec4.2.1}, where $\|A^{(1)}\|_\infty$ and $\|A^{(2)}\|_\infty$ are the size of non-zero elements in $A^{(1)}$ and $A^{(2)}$, respectively and $\|\b{\Theta}\|_0$ is the number of non-zero elements of $\b{\Theta}$.}
\label{Tab:single_a_sps}
\end{table}

\begin{figure}[!ht]
\begin{center}
\includegraphics[width=12cm, height=5cm]{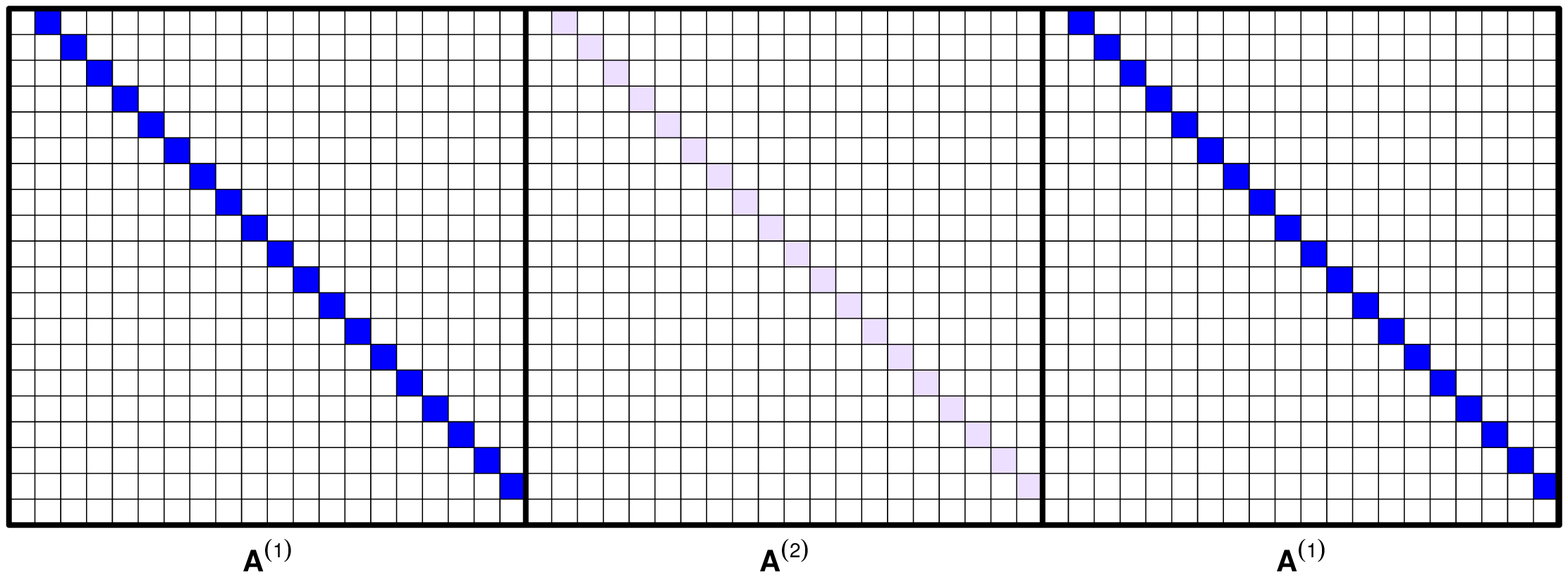}
\end{center}
\caption{The underlying coefficient matrices, $(A^{(1)}, A^{(2)}, A^{(1)})$, for the simulation setting in Section \ref{sec4.2.1}, where $A^{(2)}$ corresponds to an anomaly.}
\label{fig:single_a_sps}
\end{figure}

\begin{table}[!ht]
\centering
\begin{tabular}{ccccc}
  \hline
 & & & $A^{(1)}$ is known & $A^{(1)}$ is estimated\\
 \hline
 \multirow{4}{*}{case 1} & random  & OLS  &  100 &   90 \\ 
            &  (s = 499) & Lasso &  100 & \textbf{100}  \\   
            \cline{2-5}  
            & deterministic  & OLS  & 100 &  95 \\ 
            &  (s =  499) & Lasso &  100 &  \textbf{100}\\   
            \hline
 \multirow{4}{*}{case 2} & random  & OLS  &  100 &   30 \\ 
            &  (s = 499) & Lasso &  100 & \textbf{58}  \\   
            \cline{2-5}  
            & deterministic  & OLS  & 100 &  50 \\ 
            &  (s =  499) & Lasso &  100 &  \textbf{92}\\   
            \hline
\end{tabular}
\caption{Empirical power ($\%$) from 100 simulation runs for two methods in all cases described in Section \ref{sec4.2.1}, where $s$ is the number of intervals examined.}
\label{Tab:single_a_sps_1}
\end{table}


\begin{table}[h!]
\centering
\begin{tabular}{ccccc}
  \hline
 & & & $A^{(1)}$ is known & $A^{(1)}$ is estimated\\
 \hline
 \multirow{4}{*}{case 1} & random  & OLS  & 1.43 (0.53) & 6.00 (13.03) \\ 
            &  (s = 499) & Lasso & 1.31 (0.45) & 1.28 (0.21) \\   
            \cline{2-5}  
            & deterministic  & OLS  & 0.43 (0.24) & 2.70 (9.71)\\ 
            &  (s = 499) & Lasso & 0.34 (0.12) &  0.37 (0.13)\\   
            \hline
 \multirow{4}{*}{case 2} & random  & OLS  & 3.26 (1.19) & 32.80 (20.89) \\ 
            &  (s = 499) & Lasso & 3.21 (1.18) & 20.12 (22.48) \\   
            \cline{2-5}  
            & deterministic  & OLS  & 0.43 (0.56) &  23.39 (23.12)\\ 
            &  (s = 499) & Lasso & 0.30 (0.19) &  4.03 (12.56)\\   
            \hline
\end{tabular}
\caption{The mean (standard deviation) of Hausdorff distance from 100 simulation runs for two methods in all cases described in Section \ref{sec4.2.1}, where $s$ is the number of intervals examined.}
\label{Tab:single_a_sps_2}
\end{table}

We now consider the case when $A^{(1)}$ is sparse i.e. only a smaller number of entries are non-zero. 
Similar to the settings used in \citet{safikhani2020joint}, the 1-off diagonal values of the coefficient matrix are non-zero as shown in Figure \ref{fig:single_a_sps}.
The details of the simulation setting are given in Table \ref{Tab:single_a_sps}.
Tables \ref{Tab:single_a_sps_1} and \ref{Tab:single_a_sps_2} show similar interpretations with those given in Section \ref{denseA1}.

\section{Data analysis} \label{sec5}
\subsection{Yellow cab demand in New York City}

\begin{figure}[ht!]
\begin{center}
\includegraphics[width=15cm, height=15cm]{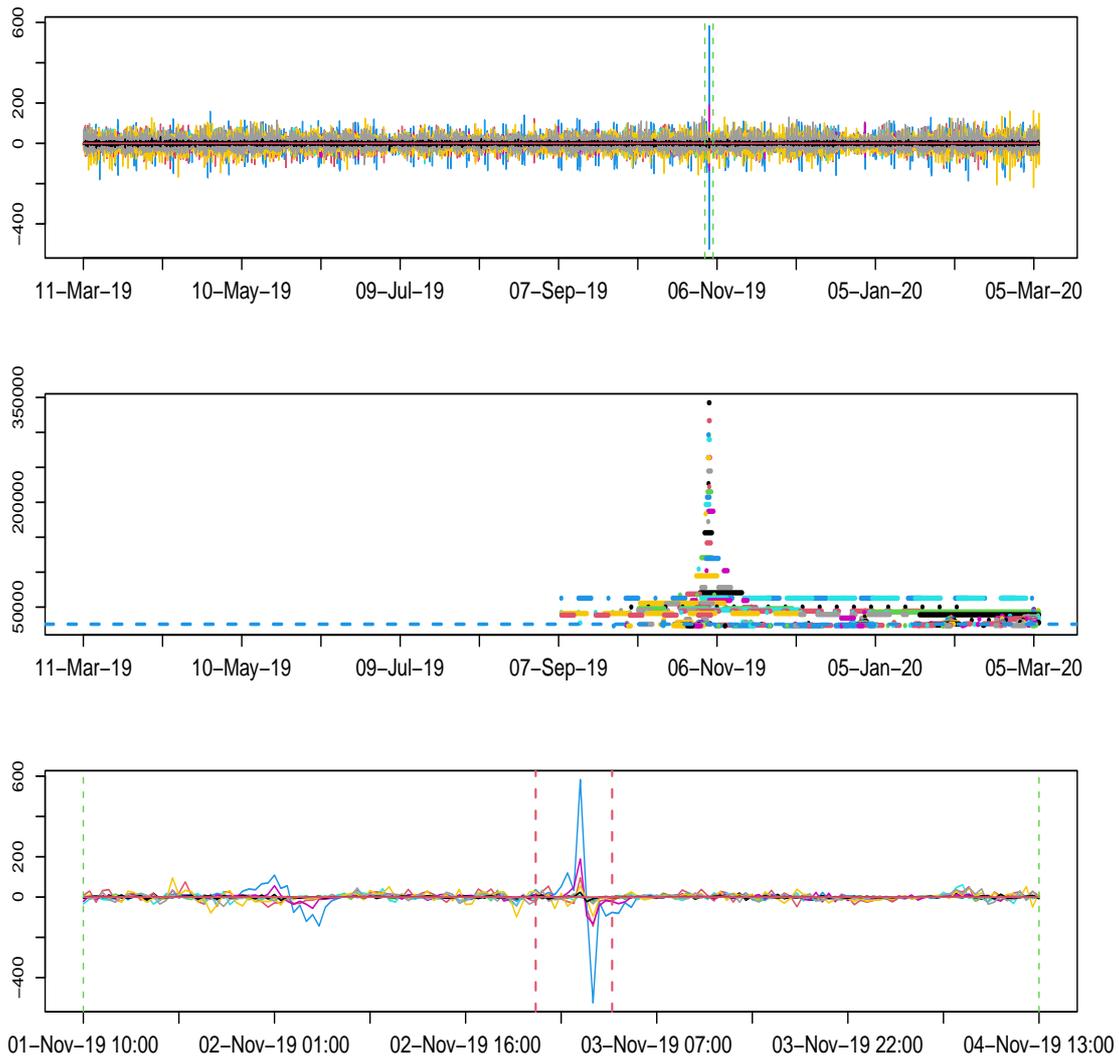}
\end{center}
 \caption{(Top) The differenced yellow taxi pickups recorded from March 11, 2019 to March 6, 2020 in Manhattan. (Middle) The 20 largest test statistics with the corresponding interval. The blue horizontal dashed line indicates the threshold. (Bottom) The portion of the top plot indicated with dashed green vertical lines. Red vertical lines show the estimated anomaly, [Nov $3, 2019$ $00:00:00$, Nov $3, 2019$ $06:30:00$].}
  \label{fig:nyc_taxi}
\end{figure}

To demonstrate the usefulness of our method, we now turn to real data applications. 
In our first example, we apply our method to the yellow taxi trip data that is previously analysed by \citet{safikhani2020joint}. 
The data can be downloaded from  the New York City Taxi and Limousine Commission (TLC) Database (\url{https://www1.nyc.gov/site/tlc/about/tlc-trip-record-data.page}). This data consists of the number of yellow taxi pick-ups recorded from 10 randomly selected zones in Manhattan, a borough in New York City.
We aggregate the number of yellow taxi pick-ups every 30 minutes from March 11, 2019 to March 6, 2020 which results in 17376 time points.
We seek to detect whether a collective anomaly exists 
after removing the first order nonstationarity from the data. We use the differenced version of the time series,
using the first 4344 data points to estimate the underlying VAR coefficient $A^{(1)}$ by applying a lasso penalty. The next 4344 data points are used to obtain a threshold, where the threshold is selected as the 99$\%$ quantile of the test statistics from 100 deterministically chosen intervals. 
Then we detect a single anomaly using the remaining 8687 data points.

\begin{figure}[ht!]
\begin{center}
\includegraphics[width=14cm, height=10cm]{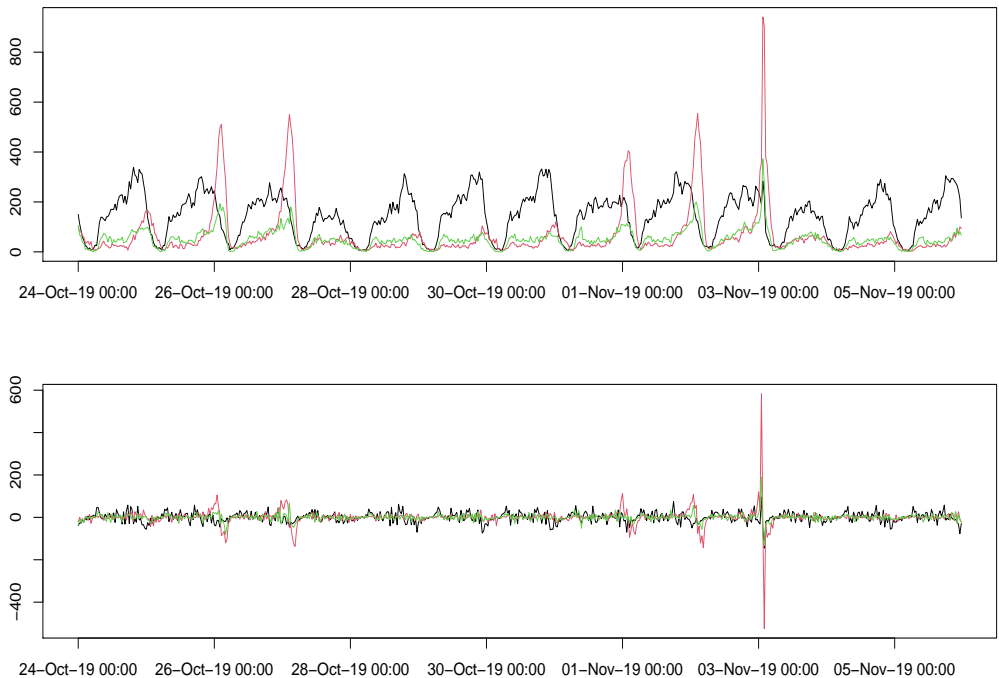}
\end{center}
 \caption{Taxi demand (Top) and differenced Taxi demand (Bottom) for 2nd (black), 4th (red) and 6th (green) zones in Manhattan recorded from October 24, 2019 to November 5, 2019.}
   \label{fig:nyc_taxi_1}
\end{figure}

The top plot in Figure \ref{fig:nyc_taxi} shows that two consecutive spikes are observed between October 7 and November 6 in 2019, where the interval within green vertical lines is enlarged in the bottom plot. 
From the middle plot, we see that the largest test statistic is obtained for a small interval which includes the spikes shown in the top plot.
The bottom plot shows that the spikes occur between 12am to 2am on November 3, 2019 and our method detects an anomaly between 12am and 6:30am on November 3, 2019.
From Figure \ref{fig:nyc_taxi_1}, we see that a sudden high demand occurred at the $4^{\text{th}}$ and $6^{\text{th}}$ zones located in Downtown Manhattan (also known as Lower Manhattan).
This anomaly seems to be related to traffic management for the 2019 New York City Marathon which took placed on November 3, 2019 in New York City. 
We can interpret that there was a  sudden high demand in Downtown Manhattan where the marathon route did not pass through, and this changes the relationship between the 10 zones we investigate.

\subsection{EEG Data}


\begin{figure}[ht!]
\begin{center}
\includegraphics[width=16cm, height=11cm]{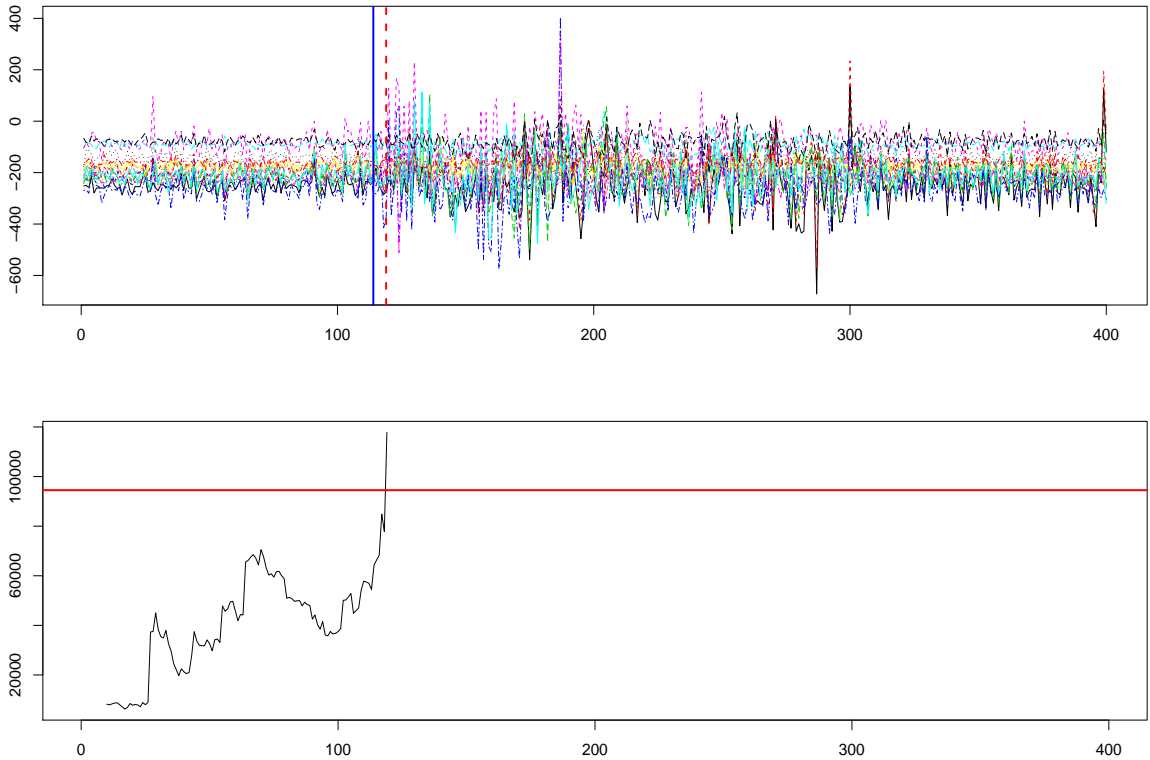}
\end{center}
 \caption{(Top) EEG data recorded at 18 different channels. Blue solid vertical line is the time at which the neurologist thinks seizure starts and the red dashed vertical line is the anomaly detected in the online setting. (Bottom) The maximum test statistics at each time point obtained through Algorithm \ref{algo_online} which stops when the anomaly is detected. The horizontal red line presents the pre-specified threshold.}
   \label{fig:eeg_online}
\end{figure}

We now show how our method can be used in as an online changepoint detection method. 
We demonstrate this on electroencephalogram (EEG) data collected from an epileptic patient. Other ways of analysing this dataset can be found in \citet{ombao2001automatic}, \citet{ombao2005slex} and \citet{schroder2019fresped}. The data consists of brain electrical potentials recorded by placing electrodes on 18 locations on the scalp of a patient. The EEG signals are recorded during an epileptic seizure, thus these exists a visible change in the data as shown in Figure \ref{fig:eeg_online}. The brain wave patterns are recorded over $500$ seconds with the sampling rate 100 Hz (i.e. 100 points per second). As done in  \citet{safikhani2020joint}, to speed up computation, we use 2 observations per second which reduces the number of time points to $T=1000$.

We separate the data into a training set of the size $T_1=600$ and a test set of the size $T_2=400$. The first half of the training set is used to estimate the underlying VAR coefficient $A^{(1)}$ by applying a lasso penalty and the second half is used to have a threshold that is chosen as the 99$\%$ quantile of the test statistics computed from 327 deterministically chosen intervals. Then we perform the single anomaly detection using a test set.

As mentioned in Section \ref{sec1}, here we show how our method can be applied to the online framework. 
We refer the reader to \citet{fisch2020real} and \citet{yu2021optimal} for the recent works on online detection algorithm for change-points or anomalies. In the online setting, we make sequential decisions about the occurrence of an anomaly whenever each new observations is obtained. 
Our algorithm for online anomaly detection is similar to Algorithm 2  of \citet{yu2021optimal}.
The detailed procedure is given in Algorithm \ref{algo_online} where we set $t_0=10$.
As shown in Figure \ref{fig:eeg_online}, an anomaly is estimated at $t=119$ that has the detection delay of $5$ time points compared to $t=114$ at which the neurologist states that a seizure takes place.

\begin{center}
\begin{algorithm}[H]
\setstretch{1}
\SetAlgoLined
\textbf{INPUT}: $\b{X}$, $\lambda^{\text{thr}}$, $t_0$ \\
 $t \leftarrow t_0$ \\
 FLAG $\leftarrow 0$\\
\While{\normalfont FLAG $= 0$}{
 $t \leftarrow t+1$ \\
 $J \leftarrow \floor{\frac{\log t}{\log 2}}$ \\
$ j \leftarrow 1$\\
  \While{\normalfont FLAG = 0 and $j \leq J$}{
  $s_j \leftarrow t - 2^{j-1} $\\
  $J \leftarrow [s_j, t] $\\
  FLAG $\leftarrow \mathbb{1} \{ T^{\text{lasso}}(J) > \lambda^\textsuperscript{thr} \}$\\
  $j \leftarrow j+1$\\
  }
 }
\textbf{OUTPUT} : $t$.
 \caption{Online anomaly detection}
 \label{algo_online}
\end{algorithm} 
\end{center}

\section{Discussion} \label{sec6}
Our lasso-based approach is motivated for data where we have substantially more data about the current or normal behaviour of the time series than for any anomaly or epidemic change. Thus it is natural to model the change as sparse and thus a lasso-based test is more appropriate than a standard likelihood-ratio or OLS-based test.
We provide a numerical evidence that our method outperforms existing competitors in detecting sparse change when $A^{(1)}$ is either dense or sparse.
Our method searches a set of local segments to detect an anomalous interval, whereas the existing change detection methodologies for the VAR model perform global optimisation.
As illustrated in real data example, the local optimisation aspect of our method gives a flexibility to extend it to the online setting. 



\section*{Acknowledgements}
The authors thank Hernando Ombao and Abolfazl Safikhani for kindly providing access to the EEG data, and also thank Yi Yu for helpful conversations.
The authors gratefully acknowledge the financial support of EPSRC, grant EP/N031938/1 (STATSCALE).


\bibliographystyle{apalike}
\bibliography{varref}

\begin{thebibliography}{}

\bibitem[Ansley and Kohn, 1986]{ansley1986note}
Ansley, C.~F. and Kohn, R. (1986).
\newblock A note on reparameterizing a vector autoregressive moving average
  model to enforce stationarity.
\newblock {\em Journal of Statistical Computation and Simulation}, 24:99--106.

\bibitem[Aue et~al., 2009]{aue2009break}
Aue, A., H{\"o}rmann, S., Horv{\'a}th, L., and Reimherr, M. (2009).
\newblock Break detection in the covariance structure of multivariate time
  series models.
\newblock {\em The Annals of Statistics}, 37:4046--4087.

\bibitem[Aue et~al., 2006]{aue2006change}
Aue, A., Horv{\'a}th, L., Hu{\v{s}}kov{\'a}, M., and Kokoszka, P. (2006).
\newblock Change-point monitoring in linear models.
\newblock {\em The Econometrics Journal}, 9:373--403.

\bibitem[Bai et~al., 2020]{bai2020multiple}
Bai, P., Safikhani, A., and Michailidis, G. (2020).
\newblock Multiple change points detection in low rank and sparse high
  dimensional vector autoregressive models.
\newblock {\em IEEE Transactions on Signal Processing}, 68:3074--3089.

\bibitem[Baltagi et~al., 2017]{baltagi2017identification}
Baltagi, B.~H., Kao, C., and Wang, F. (2017).
\newblock Identification and estimation of a large factor model with structural
  instability.
\newblock {\em Journal of Econometrics}, 197:87--100.

\bibitem[Ba{\'n}bura et~al., 2010]{banbura2010large}
Ba{\'n}bura, M., Giannone, D., and Reichlin, L. (2010).
\newblock Large {B}ayesian vector auto regressions.
\newblock {\em Journal of applied Econometrics}, 25:71--92.

\bibitem[Baranowski et~al., 2019]{baranowski2019narrowest}
Baranowski, R., Chen, Y., and Fryzlewicz, P. (2019).
\newblock Narrowest-over-threshold detection of multiple change points and
  change-point-like features.
\newblock {\em Journal of the Royal Statistical Society: Series B (Statistical
  Methodology)}, 81:649--672.

\bibitem[Barigozzi et~al., 2018]{barigozzi2018simultaneous}
Barigozzi, M., Cho, H., and Fryzlewicz, P. (2018).
\newblock Simultaneous multiple change-point and factor analysis for
  high-dimensional time series.
\newblock {\em Journal of Econometrics}, 206:187--225.

\bibitem[Barigozzi and Hallin, 2017]{barigozzi2017network}
Barigozzi, M. and Hallin, M. (2017).
\newblock A network analysis of the volatility of high dimensional financial
  series.
\newblock {\em Journal of the Royal Statistical Society: Series C (Applied
  Statistics)}, 66:581--605.

\bibitem[Basu et~al., 2019]{basu2019low}
Basu, S., Li, X., and Michailidis, G. (2019).
\newblock Low rank and structured modeling of high-dimensional vector
  autoregressions.
\newblock {\em IEEE Transactions on Signal Processing}, 67:1207--1222.

\bibitem[Basu and Michailidis, 2015]{basu2015regularized}
Basu, S. and Michailidis, G. (2015).
\newblock Regularized estimation in sparse high-dimensional time series models.
\newblock {\em The Annals of Statistics}, 43:1535--1567.

\bibitem[Bernanke et~al., 2005]{bernanke2005measuring}
Bernanke, B.~S., Boivin, J., and Eliasz, P. (2005).
\newblock Measuring the effects of monetary policy: a factor-augmented vector
  autoregressive (favar) approach.
\newblock {\em The Quarterly journal of economics}, 120:387--422.

\bibitem[Breitung and Eickmeier, 2011]{breitung2011testing}
Breitung, J. and Eickmeier, S. (2011).
\newblock Testing for structural breaks in dynamic factor models.
\newblock {\em Journal of Econometrics}, 163:71--84.

\bibitem[Calomiris et~al., 2008]{calomiris2008foreclosure}
Calomiris, C.~W., Longhofer, S.~D., and Miles, W. (2008).
\newblock The foreclosure-house price nexus: lessons from the 2007-2008 housing
  turmoil.
\newblock Technical report, National Bureau of Economic Research.

\bibitem[Chen and Gupta, 1997]{chen1997testing}
Chen, J. and Gupta, A.~K. (1997).
\newblock Testing and locating variance changepoints with application to stock
  prices.
\newblock {\em Journal of the American Statistical association}, 92:739--747.

\bibitem[Chen et~al., 2014]{chen2014detecting}
Chen, L., Dolado, J.~J., and Gonzalo, J. (2014).
\newblock Detecting big structural breaks in large factor models.
\newblock {\em Journal of Econometrics}, 180:30--48.

\bibitem[Chen and Tian, 2010]{chen2010modified}
Chen, Z. and Tian, Z. (2010).
\newblock Modified procedures for change point monitoring in linear models.
\newblock {\em Mathematics and computers in simulation}, 81:62--75.

\bibitem[Cho, 2016]{cho2016change}
Cho, H. (2016).
\newblock Change-point detection in panel data via double cusum statistic.
\newblock {\em Electronic Journal of Statistics}, 10:2000--2038.

\bibitem[Cho and Fryzlewicz, 2015]{cho2015multiple}
Cho, H. and Fryzlewicz, P. (2015).
\newblock Multiple-change-point detection for high dimensional time series via
  sparsified binary segmentation.
\newblock {\em Journal of the Royal Statistical Society: Series B (Statistical
  Methodology)}, 77:475--507.

\bibitem[Cribben and Yu, 2017]{cribben2017estimating}
Cribben, I. and Yu, Y. (2017).
\newblock Estimating whole-brain dynamics by using spectral clustering.
\newblock {\em Journal of the Royal Statistical Society: Series C (Applied
  Statistics)}, 66:607--627.

\bibitem[Davis et~al., 2016]{davis2016sparse}
Davis, R.~A., Zang, P., and Zheng, T. (2016).
\newblock Sparse vector autoregressive modeling.
\newblock {\em Journal of Computational and Graphical Statistics},
  25:1077--1096.

\bibitem[De~Mol et~al., 2008]{de2008forecasting}
De~Mol, C., Giannone, D., and Reichlin, L. (2008).
\newblock Forecasting using a large number of predictors: Is {B}ayesian
  shrinkage a valid alternative to principal components?
\newblock {\em Journal of Econometrics}, 146:318--328.

\bibitem[Dette and G{\"o}smann, 2020]{dette2020likelihood}
Dette, H. and G{\"o}smann, J. (2020).
\newblock A likelihood ratio approach to sequential change point detection for
  a general class of parameters.
\newblock {\em Journal of the American Statistical Association},
  115:1361--1377.

\bibitem[Fisch et~al., 2020]{fisch2020real}
Fisch, A., Bardwell, L., and Eckley, I.~A. (2020).
\newblock Real time anomaly detection and categorisation.
\newblock {\em arXiv preprint arXiv:2009.06670}.

\bibitem[Fisch et~al., 2018]{fisch2018linear}
Fisch, A. T.~M., Eckley, I.~A., and Fearnhead, P. (2018).
\newblock A linear time method for the detection of point and collective
  anomalies.
\newblock {\em arXiv preprint arXiv:1806.01947}.

\bibitem[Fremdt, 2015]{fremdt2015page}
Fremdt, S. (2015).
\newblock Page's sequential procedure for change-point detection in time series
  regression.
\newblock {\em Statistics}, 49:128--155.

\bibitem[Fryzlewicz, 2014]{fryzlewicz2014wild}
Fryzlewicz, P. (2014).
\newblock Wild binary segmentation for multiple change-point detection.
\newblock {\em The Annals of Statistics}, 42:2243--2281.

\bibitem[Horv{\'a}th et~al., 2004]{horvath2004monitoring}
Horv{\'a}th, L., Hu{\v{s}}kov{\'a}, M., Kokoszka, P., and Steinebach, J.
  (2004).
\newblock Monitoring changes in linear models.
\newblock {\em Journal of Statistical Planning and Inference}, 126:225--251.

\bibitem[Inclan and Tiao, 1994]{inclan1994use}
Inclan, C. and Tiao, G.~C. (1994).
\newblock Use of cumulative sums of squares for retrospective detection of
  changes of variance.
\newblock {\em Journal of the American Statistical Association}, 89:913--923.

\bibitem[Kim and Siegmund, 1989]{kim1989likelihood}
Kim, H.-J. and Siegmund, D. (1989).
\newblock The likelihood ratio test for a change-point in simple linear
  regression.
\newblock {\em Biometrika}, 76:409--423.

\bibitem[Kirch et~al., 2015]{kirch2015detection}
Kirch, C., Muhsal, B., and Ombao, H. (2015).
\newblock Detection of changes in multivariate time series with application to
  eeg data.
\newblock {\em Journal of the American Statistical Association},
  110:1197--1216.

\bibitem[Kov{\'a}cs et~al., 2020]{kovacs2020seeded}
Kov{\'a}cs, S., Li, H., B{\"u}hlmann, P., and Munk, A. (2020).
\newblock Seeded binary segmentation: A general methodology for fast and
  optimal change point detection.
\newblock {\em arXiv preprint arXiv:2002.06633}.

\bibitem[Lai, 1995]{lai1995sequential}
Lai, T.~L. (1995).
\newblock Sequential changepoint detection in quality control and dynamical
  systems.
\newblock {\em Journal of the Royal Statistical Society: Series B
  (Methodological)}, 57(4):613--644.

\bibitem[Lin and Michailidis, 2017]{lin2017regularized}
Lin, J. and Michailidis, G. (2017).
\newblock Regularized estimation and testing for high-dimensional multi-block
  vector-autoregressive models.
\newblock {\em The Journal of Machine Learning Research}, 18:4188--4236.

\bibitem[L{\"u}tkepohl, 2005]{lutkepohl2005new}
L{\"u}tkepohl, H. (2005).
\newblock {\em New introduction to multiple time series analysis}.
\newblock Springer Science \& Business Media.

\bibitem[Nicholson et~al., 2020]{nicholson2020high}
Nicholson, W.~B., Wilms, I., Bien, J., and Matteson, D.~S. (2020).
\newblock High dimensional forecasting via interpretable vector autoregression.
\newblock {\em Journal of Machine Learning Research}, 21:1--52.

\bibitem[Ombao et~al., 2005]{ombao2005slex}
Ombao, H., Von~Sachs, R., and Guo, W. (2005).
\newblock Slex analysis of multivariate nonstationary time series.
\newblock {\em Journal of the American Statistical Association}, 100:519--531.

\bibitem[Ombao et~al., 2001]{ombao2001automatic}
Ombao, H.~C., Raz, J.~A., von Sachs, R., and Malow, B.~A. (2001).
\newblock Automatic statistical analysis of bivariate nonstationary time
  series.
\newblock {\em Journal of the American Statistical Association}, 96:543--560.

\bibitem[Rapach et~al., 2007]{rapach2007forecasting}
Rapach, D.~E., Strauss, J.~K., et~al. (2007).
\newblock Forecasting real housing price growth in the eighth district states.
\newblock {\em Federal Reserve Bank of St. Louis. Regional Economic
  Development}, 3:33--42.

\bibitem[Safikhani and Shojaie, 2020]{safikhani2020joint}
Safikhani, A. and Shojaie, A. (2020).
\newblock Joint structural break detection and parameter estimation in
  high-dimensional nonstationary var models.
\newblock {\em Journal of the American Statistical Association}, pages 1--14.

\bibitem[Schr{\"o}der and Ombao, 2019]{schroder2019fresped}
Schr{\"o}der, A.~L. and Ombao, H. (2019).
\newblock Fresped: Frequency-specific change-point detection in epileptic
  seizure multi-channel eeg data.
\newblock {\em Journal of the American Statistical Association}, 114:115--128.

\bibitem[Seth et~al., 2015]{seth2015granger}
Seth, A.~K., Barrett, A.~B., and Barnett, L. (2015).
\newblock Granger causality analysis in neuroscience and neuroimaging.
\newblock {\em Journal of Neuroscience}, 35:3293--3297.

\bibitem[Shojaie and Michailidis, 2010]{shojaie2010discovering}
Shojaie, A. and Michailidis, G. (2010).
\newblock Discovering graphical granger causality using the truncating lasso
  penalty.
\newblock {\em Bioinformatics}, 26:i517--i523.

\bibitem[Siegmund and Venkatraman, 1995]{siegmund1995using}
Siegmund, D. and Venkatraman, E. (1995).
\newblock Using the generalized likelihood ratio statistic for sequential
  detection of a change-point.
\newblock {\em The Annals of Statistics}, pages 255--271.

\bibitem[Sims, 1980]{sims1980macroeconomics}
Sims, C.~A. (1980).
\newblock Macroeconomics and reality.
\newblock {\em Econometrica: journal of the Econometric Society}, pages 1--48.

\bibitem[Smith, 2012]{smith2012future}
Smith, S.~M. (2012).
\newblock The future of fmri connectivity.
\newblock {\em Neuroimage}, 62:1257--1266.

\bibitem[Song and Bickel, 2011]{song2011large}
Song, S. and Bickel, P.~J. (2011).
\newblock Large vector auto regressions.
\newblock {\em arXiv preprint arXiv:1106.3915}.

\bibitem[Stock and Watson, 2008]{stock2008evolution}
Stock, J.~H. and Watson, M. (2008).
\newblock The evolution of national and regional factors in us housing
  construction.
\newblock {\em Volatility and Time Series Econometrics}.

\bibitem[Tveten et~al., 2020]{tveten2020scalable}
Tveten, M., Eckley, I.~A., and Fearnhead, P. (2020).
\newblock Scalable changepoint and anomaly detection in cross-correlated data
  with an application to condition monitoring.
\newblock {\em arXiv preprint arXiv:2010.06937}.

\bibitem[Wang et~al., 2017]{wang2017optimal}
Wang, D., Yu, Y., and Rinaldo, A. (2017).
\newblock Optimal covariance change point localization in high dimension.
\newblock {\em arXiv preprint arXiv:1712.09912}.

\bibitem[Wang et~al., 2019]{wang2019localizing}
Wang, D., Yu, Y., Rinaldo, A., and Willett, R. (2019).
\newblock Localizing changes in high-dimensional vector autoregressive
  processes.
\newblock {\em arXiv preprint arXiv:1909.06359}.

\bibitem[Wang and Samworth, 2018]{wang2018high}
Wang, T. and Samworth, R.~J. (2018).
\newblock High dimensional change point estimation via sparse projection.
\newblock {\em Journal of the Royal Statistical Society: Series B (Statistical
  Methodology)}, 80:57--83.

\bibitem[Yao, 1993]{yao1993tests}
Yao, Q. (1993).
\newblock Tests for change-points with epidemic alternatives.
\newblock {\em Biometrika}, 80(1):179--191.

\bibitem[Yau and Zhao, 2016]{yau2016inference}
Yau, C.~Y. and Zhao, Z. (2016).
\newblock Inference for multiple change points in time series via likelihood
  ratio scan statistics.
\newblock {\em Journal of the Royal Statistical Society: Series B: Statistical
  Methodology}, pages 895--916.

\bibitem[Yu et~al., 2021]{yu2021optimal}
Yu, Y., Padilla, O. H.~M., Wang, D., and Rinaldo, A. (2021).
\newblock Optimal network online change point localisation.
\newblock {\em arXiv preprint arXiv:2101.05477}.

\end{thebibliography}

\newpage
\pagenumbering{arabic}
\setcounter{page}{1}

\title{\bf Supplementary Material for Collective anomaly detection in High-dimensional VAR Models}

\author{Hyeyoung Maeng, Idris Eckley and Paul Fearnhead \hspace{.2cm} \\
    Lancaster University, United Kingdom}
  \maketitle

\setcounter{section}{0}
\section{Technical proofs} \label{pf}
We first give a preparatory lemma and then move onto the proofs of main theorems and corollaries presented in Section \ref{sec3}.

\begin{Lem} \label{lem1-1}
Let Assumptions \ref{asmpt1}-\ref{asmpt4} hold. For any $J \in \mathbb{J}_{T, p} (L)$ such that $J \subseteq [\eta_1, \eta_2]$ and any $|J| \succsim c^{\b{\mathfrak{m}}, \mathcal{M}}_1 \log p$, with probability at least $1-T^{-6}$, we have
\begin{equation*}
 {\b{\Theta}}^\top \b{X}_{J}^\top \b{X}_{J} {\b{\Theta}} \geq c^{\b{\mathfrak{m}}, \mathcal{M}}_2 |J| \cdot \|{\b{\Theta}}\|^2_2 - c^{\b{\mathfrak{m}}, \mathcal{M}}_3 \log(p)  \cdot \|{\b{\Theta}}\|^2_1
\end{equation*}
where 
$c^{\b{\mathfrak{m}}, \mathcal{M}}_1, c^{\b{\mathfrak{m}}, \mathcal{M}}_2, c^{\b{\mathfrak{m}}, \mathcal{M}}_3 >0$ are some constants depending on $\b{\mathfrak{m}}$ and $\mathcal{M}$.
\end{Lem}

\paragraph{Proof of Lemma \ref{lem1-1}}
The argument follows the proof of Lemma 13-(b) of \citet{wang2019localizing}.

\paragraph{Proof of Theorem \ref{theorem1}}
By the construction of the candidate set $\mathbb{I}^*$, it is sufficient to show that $T^{\text{lasso}}(J) \rightarrow 0$ under the null, where $T^{\text{lasso}}(J)$ is as in \eqref{lasso}.
The KKT conditions for the lasso problem in \eqref{lasso} is that any $\hat{\b{\Theta}}$ is optimal if and only if there exists a subgradient $\hat{s}$ such that 
\begin{equation} \label{kkt}
\b{X}_{J}^\top \Big(\b{Y}_{J}-\b{X}_{J}\hat{\b{\Theta}}\Big) = \lambda \hat{s}_{J},
\end{equation}
where $\hat{s}_{J} = \partial\big|\hat{\b{\Theta}}\big|_1$ is a subgradient of the $l_1$ norm evaluated at $\hat{\b{\Theta}}$ which takes the form
\begin{equation} \label{subgrad}
\hat{s}_{J} = \text{sgn}(\hat{\b{\Theta}}) \; \text{for} \; \hat{\b{\Theta}} \neq 0,  \quad |\hat{s}_{J}| \leq 1 \; \text{otherwise}.
\end{equation}
As $\b{Y}=\b{E}$ under the null, \eqref{kkt} and \eqref{subgrad} give a condition on $\b{X}$ and $\b{E}$ to ensure that we estimate $\b{\Theta}=\b{0}$ as follows: for any $J \in \mathbb{J}_{T, p} (L)$ such that $J \cap [\eta_1, \eta_2] = \emptyset$,
\begin{equation} \label{toshow}
\max_{J: J \in \mathbb{J}_{T, p} (L), J \cap [\eta_1, \eta_2] = \emptyset} \Big\|\b{X}_{J}^\top\b{E}_{J}\Big\|_\infty \leq \lambda.
\end{equation}
We remind that $\mathcal{X}_{J}$ is the unvectorised covariates as follows:
\begin{equation*}
\mathcal{X}_{J} =\begin{pmatrix}
 0 \\
 0\\
\vdots   \\
 0  \\
\b{x}_{t+1}^\prime   \\
 \vdots   \\
 \b{x}_{t+h-1}^\prime 
\end{pmatrix}_{(2h-1) \times p}
\end{equation*}
and note that
\begin{align*}
\max_{J: J \in \mathbb{J}_{T, p} (L), J \cap [\eta_1, \eta_2] = \emptyset} \Bigg\|\frac{\b{X}_{J}^\top\b{E}_{J}}{|J|}\Bigg\|_\infty & = \max_{\{J: J \in \mathbb{J}_{T, p} (L), J \cap [\eta_1, \eta_2] = \emptyset \}, 1 \leq i, j \leq p} \Bigg| e^{\prime}_i \bigg( \frac{\mathcal{X}^{\prime}_{J} E_{J}}{|J|}\bigg) e_j\Bigg|, \\ \numberthis \label{choices} 
&\leq \max_{\{J: J \in \mathbb{J}_{T, p} (L), J \cap [\eta_1, \eta_2] = \emptyset \}, 1 \leq i, j \leq p} \Bigg| e^{\prime}_i \bigg( \frac{\mathcal{X}^{\prime}_{J} E_{J}}{L}\bigg) e_j\Bigg|, 
\end{align*}
where $e_i \in \mathbb{R}^p$ with the $i$-th element equals to $1$ and zero otherwise. Similar to the argument used in Proposition $2.4(b)$  of \citet{basu2015regularized}, for fixed $i, j, J$, there exist $k_1, k_2 >0$ such that for all $\gamma >0$:
\begin{equation} \label{bm}
P\Bigg(\bigg| e^{\prime}_i \bigg( \mathcal{X}^{\prime}_{J} E_{J}\bigg) e_j\bigg| > k_1 L \gamma \Bigg) \leq 6 \exp (-k_2 L \min (\gamma, \gamma^2)).
\end{equation}
As the number of intervals contained in $\mathbb{J}_{T, p} (L)$ is of the order $O(T)$ when they are constructed through the seeded interval idea in \citet{kovacs2020seeded}, we consider the union over $p^2\cdot T$ possible choices of $i, j, J$ in \eqref{choices}.  Then the result follows by setting $\gamma =  k_3 \sqrt{\frac{2\log{p} + \log{T}}{L}}$ for a large enough $k_3 > 0$. 
Therefore, with probability at least $1-C_4 \exp(-C_5 (2\log{p} + \log{T}))$, we have 
\begin{equation} \label{ub}
\max_{J: J \in \mathbb{J}_{T, p} (L), J \cap [\eta_1, \eta_2] = \emptyset} \Big\|\b{X}_{J}^\top\b{E}_{J}\Big\|_\infty \leq C_3  \sqrt{L \log({T \lor p})},
\end{equation}
where $C_4>0$ and $C_5>0$. 
Having the condition $\lambda = C_3 \sqrt{L \log({T \lor p})}$ with a large enough $C_3 > 0$ in \eqref{toshow}, we obtain $\hat{\b{\Theta}}=\b{0}$ with probability at least $1-C_4 \exp(-C_5 (2\log{p} + \log{T}))$.
Therefore, under the null, the probability that $T^{\text{lasso}}(J) \rightarrow 0$ for $J \in \mathbb{J}_{T, p} (L)$ such that $ J \cap [\eta_1, \eta_2] = \emptyset$ is at least  $1-C_4 \exp(-C_5 (2\log{p} + \log{T}))$ where $C_4, C_5 >0$.

We emphasise that \eqref{ub} can be applied to any serially uncorrelated Gaussian errors $\b{\varepsilon}_t  \stackrel{\text{i.i.d.}}{\sim} N(\b{0}, \Sigma_{\varepsilon})$ as the constant $k_1$ in \eqref{bm} presented in Proposition $2.4(b)$  of \citet{basu2015regularized} has a form of
\begin{equation*} 
k_1 = 2 \pi \Lambda_{\max}(\Sigma_{\varepsilon})  \bigg( 1+ \frac{1+\mu_{\max}(\mathcal{A})}{\mu_{\min}(\mathcal{A})}\bigg),
\end{equation*}
where $\Lambda_{\max}(\Sigma_{\varepsilon})$ is the maximum eigenvalue of $\Sigma_{\varepsilon}$, $\mu_{\max}(\mathcal{A})=\max_{|z|=1} \Lambda_{\max} (\mathcal{A}^*(z) \mathcal{A}(z))$, $\mu_{\min}(\mathcal{A})=\min_{|z|=1} \Lambda_{\min} (\mathcal{A}^*(z) \mathcal{A}(z))$ and $\mathcal{A}(z)=\mathit{I}_p - \mathbf{A}^{(1)}z$ for the VAR(1) model and $\mathcal{A}(z)=\mathit{I}_p - \sum_{d=1}^{q} \mathbf{A}^{(1)}_d z^d$ for the VAR(q) model. Therefore, even if $\Sigma_{\varepsilon}$ is not an identity matrix, we can have \eqref{ub} with a different constant $C_3$ which depends on the maximum eigenvalue of $\Sigma_{\varepsilon}$.

\paragraph{Proof of Theorem \ref{theorem2}}
It is sufficient to prove that for any $J \in \mathbb{J}_{T, p} (L)$ such that $ J \subseteq [\eta_1, \eta_2]$, with probability approaching to $1$ as $T \rightarrow \infty$,
\begin{equation} \label{toshowlem2}
\big\|\b{Y}_{J} \big\|_2^2  - \big\|\b{Y}_{J}-\b{X}_{J}{\b{\Theta}} \big\|_2^2 > \lambda\|\b{\Theta}\|_1 + \lambda^\textsuperscript{thr}.
\end{equation}
This is because the other part in equation \eqref{lasso_alt}, 
\begin{equation} \label{toshowlem2_0}
\Big\{  \big\|\b{Y}_{J}-\b{X}_{J}{\b{\Theta}} \big\|_2^2 + \lambda\|\b{\Theta}\|_1 - \big\|\b{Y}_{J}-\b{X}_{J}\hat{\b{\Theta}} \big\|_2^2 - \lambda\|\hat{\b{\Theta}}\|_1 \Big\},    
\end{equation}
is always positive and the left-hand side of \eqref{toshowlem2} dominates \eqref{toshowlem2_0}. We can simplify \eqref{toshowlem2} as
 \begin{equation}  \label{toshowlem2_1}
 {\b{\Theta}}^\top \b{X}_{J}^\top \b{X}_{J} {\b{\Theta}} + 2 {\b{\Theta}}^\top \b{X}_{J}^\top \b{E}_{J} > \lambda\|\b{\Theta}\|_1 + \lambda^\textsuperscript{thr}.
\end{equation}
The left-hand side of \eqref{toshowlem2_1} is a Gaussian variable that can be written as  $\b{\nu}_{J}^\top\b{\nu}_{J} + 2 \b{\nu}_{J}^\top\b{E}_{J} \sim N(\b{\nu}_{J}^\top\b{\nu}_{J}, 4 \b{\nu}_{J}^\top \Sigma_{\varepsilon} \b{\nu}_{J})$, where $\b{\nu}_{J} = \b{X}_{J} {\b{\Theta}}$ and $\b{\nu}_{J}^\top\b{\nu}_{J} \rightarrow \infty$. Then for any $g(J) = o(\sqrt{\b{\nu}_{J}^\top\b{\nu}_{J}})$ that goes to $\infty$, we have the following bound with probability approaching to $1$,
\begin{equation} \label{bound_vv}
\b{\nu}_{J}^\top\b{\nu}_{J} + 2 \b{\nu}_{J}^\top\b{E}_{J} \geq \b{\nu}_{J}^\top\b{\nu}_{J} - g(J) \sqrt{4 \gamma \b{\nu}_{J}^\top \b{\nu}_{J}},
\end{equation}
where $\gamma$ is the maximum eigenvalue of $\Sigma_{\varepsilon}$. The right-hand side of \eqref{bound_vv} is of order $\b{\nu}_{J}^\top\b{\nu}_{J}$, thus we now show that ${\b{\Theta}}^\top \b{X}_{J}^\top \b{X}_{J} {\b{\Theta}}$ is bounded by $\lambda\|\b{\Theta}\|_1 + \lambda^\textsuperscript{thr}$ with probability tending to $1$.
From Lemma \ref{lem1-1}, with probability approaching to $1$, we have
\begin{align*}
 {\b{\Theta}}^\top \b{X}_{J}^\top \b{X}_{J} {\b{\Theta}} &\geq c^{\b{\mathfrak{m}}, \mathcal{M}}_2 |J| \cdot \|{\b{\Theta}}\|^2_2 - c^{\b{\mathfrak{m}}, \mathcal{M}}_3 \log(p)  \cdot \|{\b{\Theta}}\|^2_1 \\
 &\geq c^{\b{\mathfrak{m}}, \mathcal{M}}_2 L \cdot \|{\b{\Theta}}\|^2_2 - c^{\b{\mathfrak{m}}, \mathcal{M}}_3 \log(p)  \cdot \|{\b{\Theta}}\|^2_1,
\end{align*}
where $c^{\b{\mathfrak{m}}, \mathcal{M}}_2, c^{\b{\mathfrak{m}}, \mathcal{M}}_3 >0$, thus we now show 
\begin{equation} \label{pf_thm2_1}
c^{\b{\mathfrak{m}}, \mathcal{M}}_2  \|{\b{\Theta}}\|^2_2 >  c^{\b{\mathfrak{m}}, \mathcal{M}}_3 \frac{\log(p)}{L} \cdot \|{\b{\Theta}}\|^2_1 + \frac{\lambda}{L}\|\b{\Theta}\|_1 + \frac{\lambda^\textsuperscript{thr}}{L},
\end{equation} 
as $T, p \rightarrow \infty$. We can obtain \eqref{pf_thm2_1} as $T, p \rightarrow \infty$ from combining 
\begin{align*}
& (a) \;c^{\b{\mathfrak{m}}, \mathcal{M}}_2  \|{\b{\Theta}}\|^2_2 >  c^{\b{\mathfrak{m}}, \mathcal{M}}_3 \frac{\log(p)}{L} \cdot \|{\b{\Theta}}\|^2_1,\\
& (b) \; c^{\b{\mathfrak{m}}, \mathcal{M}}_2  \|{\b{\Theta}}\|^2_2 > \frac{\lambda}{L}\|\b{\Theta}\|_1, \\
& (c) \; c^{\b{\mathfrak{m}}, \mathcal{M}}_2  \|{\b{\Theta}}\|^2_2 > \frac{\lambda^\textsuperscript{thr}}{L},
\end{align*}
where (a) can be shown by using $d_0 \|\b{\Theta}\|_2^2 \geq \|\b{\Theta}\|_1^2$ from Assumption \ref{asmpt3} and $\frac{\log p}{L} \rightarrow 0$ from Assumption \ref{asmpt2}. By using $d_0 \|\b{\Theta}\|_2^2 \geq \|\b{\Theta}\|_1^2$, (b) becomes $c^{\b{\mathfrak{m}}, \mathcal{M}}_2  \|{\b{\Theta}}\|_2 > \frac{\lambda}{L} \sqrt{d_0}$ that can be achieved from ${\frac{\lambda}{L}}  = \sqrt{\frac{ C_3 \log(T \lor p)}{L}}$ and $\|\b{\Theta}\|_2^2  > C_2 \cdot \frac{\log^{1+\xi}{(T \lor p)}}{L}$ in Assumption \ref{asmpt4}. Similarly (c) can be obtained from $ \frac{\lambda^\textsuperscript{thr}}{L} = O\Bigg(\sqrt{\frac{\log(T \lor p)}{L}}\Bigg)$ and  Assumption \ref{asmpt4}.

We now consider the case $\Sigma_\varepsilon$ is not an identity matrix. In that case, \eqref{toshowlem2} becomes
\begin{equation*} 
\b{Y}_{J}^\top \Sigma_{\varepsilon}^{-1} \b{Y}_{J}  - (\b{Y}_{J}-\b{X}_{J}{\b{\Theta}})^\top \Sigma_{\varepsilon}^{-1} (\b{Y}_{J}-\b{X}_{J}{\b{\Theta}}) > \lambda\|\b{\Theta}\|_1 + \lambda^\textsuperscript{thr},
\end{equation*}
thus \eqref{toshowlem2_1} becomes
 \begin{equation} \label{pf_thm2_sigma_eq1}
 {\b{\Theta}}^\top \b{X}_{J}^\top \Sigma_{\varepsilon}^{-1} \b{X}_{J} {\b{\Theta}} + 2 {\b{\Theta}}^\top \b{X}_{J}^\top\Sigma_{\varepsilon}^{-1}  \b{E}_{J} > \lambda\|\b{\Theta}\|_1 + \lambda^\textsuperscript{thr},
\end{equation}
which holds by following the same argument used above with $\b{\nu}_{J} = \Sigma_{\varepsilon}^{-1/2} \b{X}_{J} {\b{\Theta}}$ and different constants, as the left-hand side of \eqref{pf_thm2_sigma_eq1} is a Gaussian random variable bounded by a component that is of order $\b{\nu}_{J}^\top\b{\nu}_{J}$.

Lastly, without repeating all the proofs, we argue that the theory we present for the known $\Sigma_{\varepsilon}$ can be applied to the case when an estimate of ${\Sigma_{\varepsilon}}$ is used.
If $\hat{\Sigma}_{\varepsilon}$ is used instead of ${\Sigma_{\varepsilon}}$, the left-hand side of \eqref{pf_thm2_sigma_eq1} can be rewritten as
\begin{equation} \label{pf_thm2_sigma_eq3}
   {\b{\Theta}}^\top \b{X}_{J}^\top \Sigma_{\varepsilon}^{-1} \b{X}_{J} {\b{\Theta}} + 2 {\b{\Theta}}^\top \b{X}_{J}^\top\Sigma_{\varepsilon}^{-1}  \b{E}_{J}+  {\b{\Theta}}^\top \b{X}_{J}^\top (\hat{\Sigma_{\varepsilon}}^{-1}-\Sigma_{\varepsilon}^{-1}) \b{X}_{J} {\b{\Theta}} + 2 {\b{\Theta}}^\top \b{X}_{J}^\top (\hat{\Sigma_{\varepsilon}}^{-1}-\Sigma_{\varepsilon}^{-1})  \b{E}_{J},
\end{equation}
thus the test depends on the eigenvalues of the measure of the distance between $\hat{\Sigma}^{-1}_{\varepsilon}$ and $\Sigma^{-1}_{\varepsilon}$. If $\hat{\Sigma}^{-1}_{\varepsilon}$ converges to $\Sigma^{-1}_{\varepsilon}$ as observation increases, the last two terms in \eqref{pf_thm2_sigma_eq3} become under control, thus we can obtain the same argument with extra constant terms.

\paragraph{Proof of Theorem \ref{thm2-1}}
It is straightforward that the test statistic of the OLS method in \eqref{naive} has a $\chi^2_{p^2}$ distribution, where the degrees of freedom $p^2$ comes from the difference in dimensionality of ${\Theta}_0$ and $\hat{\Theta}$. 
Therefore, we get an asymptotic level $\alpha$ test if the null hypothesis is rejected for $T(J) > \chi^2_{p^2; (1-\alpha)}$, where $\chi^2_{p^2; (1-\alpha)}$ is the $(1-\alpha)$-quantile of chi-square distribution with $p^2$ degrees of freedom.
Using the threshold established above, under the alternative, an upper bound on the power of the OLS method can be obtained as
\begin{align*}
& P\bigg( T(J) > \chi^2_{p^2; (1-\alpha)} \bigg) \\ 
& = P\bigg(\|\b{Y}_{J} \|_2^2  -\|\b{Y}_{J}-\b{X}_{J}{\b{\Theta}} \|_2^2 + \Big\{ \|\b{Y}_{J}-\b{X}_{J}{\b{\Theta}} \|_2^2 -\|\b{Y}_{J}-\b{X}_{J}\hat{\b{\Theta}}\|_2^2 \Big\} \geq \chi^2_{p^2; (1-\alpha)}\bigg) \\ \numberthis \label{pf_tm3_1}
& = P\bigg(\|\b{Y}_{J} \|_2^2  -\|\b{Y}_{J}-\b{X}_{J}{\b{\Theta}} \|_2^2 + \Big\{ \|\b{Y}_{J}-\b{X}_{J}{\b{\Theta}} \|_2^2 -\|\b{Y}_{J}-\b{X}_{J}\hat{\b{\Theta}}\|_2^2 \Big\} - p^2 \geq \chi^2_{p^2; (1-\alpha)} - p^2 \bigg) \\ \numberthis \label{pf_tm3_2}
& \leq \frac{E(\|\b{Y}_{J} \|_2^2  -\|\b{Y}_{J}-\b{X}_{J}{\b{\Theta}} \|_2^2)}{\chi^2_{p^2; (1-\alpha)} - p^2} \\ \numberthis \label{pf_tm3_3}
& \approx \frac{E(\|\b{Y}_{J} \|_2^2  -\|\b{Y}_{J}-\b{X}_{J}{\b{\Theta}} \|_2^2)}{\frac{1}{2}\Big(z_{1-\alpha} + \sqrt{2p^2-1}\Big)^2 - p^2},
\end{align*}
where $z_{1-\alpha}$ is the $(1-\alpha)$-quantile of Gaussian distribution.
The equality in \eqref{pf_tm3_1} is obtained by subtracting $E\Big\{ \|\b{Y}_{J}-\b{X}_{J}{\b{\Theta}} \|_2^2 -\|\b{Y}_{J}-\b{X}_{J}\hat{\b{\Theta}}\|_2^2 \Big\} = p^2$ from both sides, the inequality in \eqref{pf_tm3_2} is obtained by using Markov's inequality and \eqref{pf_tm3_3}  is achieved as the quantile of chi-square distribution has an approximation, $\chi^2_{p^2; (1-\alpha)} \approx \frac{1}{2}\Big(z_{1-\alpha} + \sqrt{2p^2-1}\Big)^2$.
Therefore, the upper bound on the power of the OLS method can be obtained as in \eqref{upperbound}, which implies that $E(\|\b{Y} \|_2^2-\|\b{Y}-\b{X}{\b{\Theta}}\|_2^2)$ needs to be at least $O_p(p)$ to have power approaching to $1$.

\paragraph{Proof of Corollary \ref{cor1}}
As we have $\b{Y}=\b{E} +  \text{vec} \big(  \mathcal{X}^{(1)}(\b{\theta}^{(1)}-{\hat{\b{\theta}}^{(1)}}^{\prime} )\big)$ under the null rather than $\b{Y}=\b{E}$, the right-hand side of the  inequality in \eqref{choices} can be represented as
\begin{align*}
&\max_{\{J: J \in \mathbb{J}_{T, p} (L), J \cap [\eta_1, \eta_2] = \emptyset \}, 1 \leq i, j \leq p} \Bigg| e^{\prime}_i \Bigg( \frac{{\mathcal{X}^{(2)}_{J}}^{\prime} \Big(E_{J}+\mathcal{X}_{J}^{(1)}\big({\b{\theta}^{(1)}}^{\prime}-{\hat{\b{\theta}}^{(1)}}^{\prime} \big) \Big)  }{L}\Bigg) e_j\Bigg| \\ \numberthis \label{a1hat_choices_1}
& \leq \max_{\{J: J \in \mathbb{J}_{T, p} (L), J \cap [\eta_1, \eta_2] = \emptyset \}, 1 \leq i, j \leq p} \Bigg| e^{\prime}_i \Bigg( \frac{{\mathcal{X}^{(2)}_{J}}^{\prime} E_{J}  }{L}\Bigg) e_j\Bigg| + \max_{\{J: J \in \mathbb{J}_{T, p} (L), J \cap [\eta_1, \eta_2] = \emptyset \}, 1 \leq i, j \leq p} \Bigg| e^{\prime}_i \Bigg( \frac{{\mathcal{X}^{(2)}_{J}}^{\prime} \mathcal{X}_{J}^{(1)}\big({\b{\theta}^{(1)}}^{\prime}-{\hat{\b{\theta}}^{(1)}}^{\prime} \big) }{L}\Bigg) e_j\Bigg|, 
\end{align*}
It is sufficient to show that both terms in \eqref{a1hat_choices_1} are less than or equal to  $C_3 \lambda$ with probability approaching 1.  The condition for the first term is obtained from the proof of Theorem \ref{theorem1} and the one for the second term is obtained from Assumption \ref{asmpt5} and from the fact that $\frac{\mathcal{X}^{(2)\prime}_{J}\mathcal{X}_{J}^{(1)}}{|J|}$ converges as $T \rightarrow \infty$.

\paragraph{Proof of Corollary 2}
It is sufficient to prove that \eqref{toshowlem2} still holds where $\b{Y}_{J} = \text{vec} \big(\mathcal{Y}_{J} - \mathcal{X}^{(1)}_{J}{{\b{\theta}}^{(1)}}^{\prime} \big)$ is replaced by $\b{Y}_{J}^{\prime} = \text{vec} \big(\mathcal{Y}_{J} - \mathcal{X}^{(1)}_{J}{\hat{\b{\theta}}^{(1)}}^{\prime} \big)$.
The left-hand side of \eqref{toshowlem2} can be simplified as
 \begin{equation} \label{toshow2lem2_a1hat}
 {\b{\Theta}}^\top \b{X}_{J}^\top \b{X}_{J} {\b{\Theta}} + 2 {\b{\Theta}}^\top \b{X}_{J}^\top \b{E}_{J}  + 2 {\b{\Theta}}^\top \b{X}_{J}^\top \text{vec}\Big(\mathcal{X}_{J}^{(1)}\big({\b{\theta}^{(1)}}^{\prime}-{\hat{\b{\theta}}^{(1)}}^{\prime} \big) \Big) > \lambda\|\b{\Theta}\|_1 + \lambda^\textsuperscript{thr}.
\end{equation}
As shown in the proof of Theorem 2, it is sufficient to show that  
${\b{\Theta}}^\top \b{X}_{J}^\top \b{X}_{J} {\b{\Theta}}$  is bounded by $\lambda\|\b{\Theta}\|_1 + \lambda^\textsuperscript{thr}$ with probability tending to $1$ as the last component in left-hand side of \eqref{toshow2lem2_a1hat},
\begin{equation*}
2 {\b{\Theta}}^\top \b{X}_{J}^\top \text{vec}(\mathcal{X}_{J}^{(1)}\big({\b{\theta}^{(1)}}^{\prime}-{\hat{\b{\theta}}^{(1)}}^{\prime} \big)) = 2  \b{\Theta}^\top \text{vec}\Big({\mathcal{X}^{(2)}_{J}}^{\prime} \mathcal{X}_{J}^{(1)} \big({\b{\theta}^{(1)}}^{\prime}-{\hat{\b{\theta}}^{(1)}}^{\prime} \big) \Big),
\end{equation*}
is less than $\lambda\|\b{\Theta}\|_1 + \lambda^\textsuperscript{thr}$ with probability approaching to $1$ from Assumption 5 and also from the fact that $\frac{{\mathcal{X}^{(2)}_{J}}^{\prime} \mathcal{X}_{J}^{(1)}}{|J|}$ converges as $T \rightarrow \infty$. 
Following the same logic presented in the proof Theorem 2, it can be shown that the first component in left-hand side of \eqref{toshow2lem2_a1hat} is greater than $\lambda\|\b{\Theta}\|_1 + \lambda^\textsuperscript{thr}$ with probability approaching to $1$ which completes the proof.


\section{Additional Simulation Results} \label{ASR}

In this section, additional simulation results are reported. 
As mentioned in the main paper, our method is compared with the one proposed in \citet{safikhani2020joint} available from \url{https://github.com/abolfazlsafikhani/SBDetection}.
We first present a new simulation scenario that is similar to the one used in \citet{safikhani2020joint}, then give the additional results for those scenarios examined in the main paper. 
Regarding the tuning parameters for \citet{safikhani2020joint}, we follow the recommendation of their paper by using the default ones in Section \ref{sim_stronger} where the simulation setting is a slightly modified version of scenario 1 of \citet{safikhani2020joint}.
However, the anomalies presented in Section \ref{sigma_estimated} are harder to detect as the size of change in coefficient matrix is smaller and the noise has a larger variance compared to the one in Section \ref{sim_stronger}.
Thus, to improve the performance of their method, we adjust tuning parameters rather than using the default ones and the details can be found in Section \ref{sigma_estimated}

\pagebreak
\subsection{Stronger signal-to-noise ratio and larger change size} \label{sim_stronger}
\subsubsection{Sparse $A^{(1)}$} \label{sim_stronger_1}

\begin{figure}[ht!]
\begin{center}
\includegraphics[width=10cm, height=4cm]{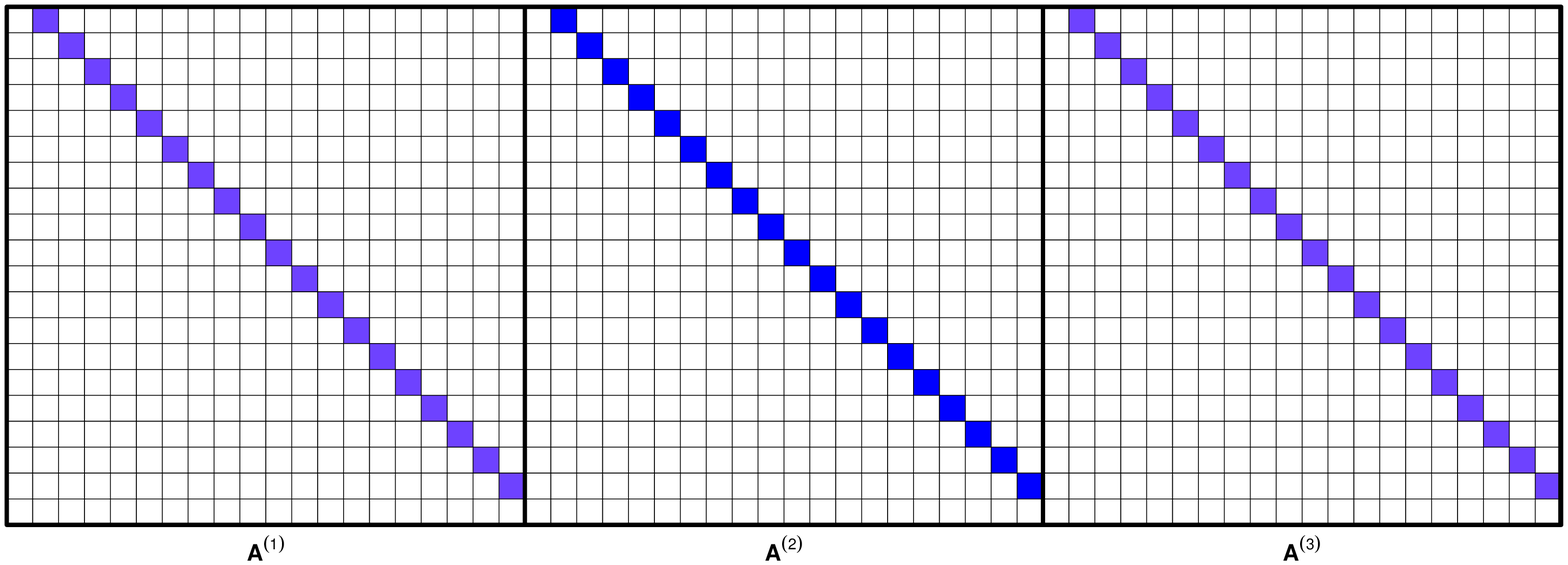}
\end{center}
\caption{The underlying coefficient matrices, $(A^{(1)}, A^{(2)}, A^{(1)})$, for the simulation setting in Section \ref{sim_stronger_1}, where $A^{(2)}$ corresponds to an anomaly.}
\label{fig:sbd_A1}
\end{figure}

\begin{table}[!ht] 
\centering
\begin{tabular}{cccccccc} 
  \hline
T & p &  $[\eta_1, \eta_2]$ & $\eta_2-\eta_1$ & $\text{nzr}(A^{(1)})$ & $\text{nzr}(A^{(2)})$  & $\|\b{\Theta}\|_0$ & $\Sigma_{\varepsilon}$ \\
  \hline
500 & 20 & $[T(1/3), T(2/3)]$ & $167$ & $-0.6$ & $0.75$ & $19$ & $0.01\textbf{\textit{I}}_p$\\
  \hline
\end{tabular} 
\caption{Simulation setting for Section \ref{sim_stronger_1}, where $\text{nzr}(A^{(1)})$ and $\text{nzr}(A^{(2)})$ are the non-zero elements of $A^{(1)}$ and $A^{(2)}$, respectively and $\|\b{\Theta}\|_0$ is the number of non-zero elements of $\b{\Theta}$.}
\label{Tab:stronger_setting}
\end{table}

We borrow the simulation setting of scenario 1 used in \citet{safikhani2020joint}.
To make the single anomaly setting, we slightly modify the original setting by changing the size of non-zero coefficients to $(-0.6, 0.75, -0.6)$ for those intervals divided by an anomaly, whereas \citet{safikhani2020joint} consider the two change points with the corresponding size of non-zero coefficients $(-0.6, 0.75, -0.8)$.
The details of the anomaly are given in Table \ref{Tab:stronger_setting}, and we can see that the size of change is larger, the length of anomaly is longer and the signal-to-noise ratio is larger than the simulation setting in Section \ref{sec4.2.1}.


\begin{table}[h!]
\centering
\begin{tabular}{ccccc}
  \hline
 & & & $A^{(1)}$ is known & $A^{(1)}$ is estimated\\
 \hline
 \multirow{4}{*}{$\Sigma_{\varepsilon}$ is known} & random  & OLS  &  100 & 100  \\   
            &  (s = 499) & Lasso &  100 & 100  \\   
            \cline{2-5}  
            & deterministic  & OLS  & 100 & 100  \\   
            &  (s = 499) & Lasso &  100 & 100  \\    
            \hline
\multirow{4}{*}{$\Sigma_{\varepsilon}$ is unknown} & random  & OLS  & 100 & 100  \\   
            &  (s = 469) & Lasso &  100 & 100  \\   
            \cline{2-5}  
            & deterministic  & OLS  & 100 & 100  \\   
            &  (s = 469) & Lasso &  100 & 100  \\   
            \hline            
\end{tabular}
\caption{Empirical power ($\%$) from 100 simulation runs for the settings described in Section \ref{sim_stronger_1}, where $s$ is the number of intervals examined.}
\label{Tab:stronger_setting_ours1}
\end{table}

Comparing the simulation results of ours with \citet{safikhani2020joint}, Tables \ref{Tab:stronger_setting_ours1}-\ref{Tab:stronger_setting_ss} show that all methods detect one anomaly in all $100$ runs (this is shown as ``one'' anomaly for the default and the lasso methods and ``two'' change-points for the method of \citet{safikhani2020joint}).
In terms of the localisation, the default and the lasso methods work better than \citet{safikhani2020joint} as they have smaller mean and sd of Hausdorff distance. 
We emphasise that the anomaly presented in this section is easier to detect than those used in the main paper in the sense that the size of change in coefficient matrix is larger, the width of anomaly is longer and the noise variance is smaller. 


\begin{table}[h!]
\centering
\begin{tabular}{ccccc}
  \hline
 & & & $A^{(1)}$ is known & $A^{(1)}$ is estimated\\
 \hline
 \multirow{4}{*}{$\Sigma_{\varepsilon}$ is known} & random  & OLS  &  1.07 (0.10)  &  0.80 (0.00) \\ 
            &  (s = 499) & Lasso &  1.08 (0.10) & 0.80 (0.00) \\   
            \cline{2-5}  
            & deterministic  & OLS  & 0.40 (0.02) & 0.40 (0.03) \\ 
            &  (s = 499) & Lasso &  0.40 (0.02) & 0.39 (0.04)\\   
            \hline
\multirow{4}{*}{$\Sigma_{\varepsilon}$ is unknown} & random  & OLS  &  1.14 (0.09) &  0.80 (0.00) \\ 
            &  (s = 469) & Lasso &  1.03 (0.08) & 0.80 (0.00)  \\   
            \cline{2-5}  
            & deterministic  & OLS  & 0.39 (0.03) & 0.40 (0.03) \\ 
            &  (s = 469) & Lasso &  0.39 (0.03) & 0.39 (0.05)\\   
            \hline            
\end{tabular}
\caption{The mean (standard deviation) of Hausdorff distance from 100 simulation runs for the settings described in Section \ref{sim_stronger_1}, where $s$ is the number of intervals examined.}
\label{Tab:stronger_setting_ours2}
\end{table}

\begin{table}[h!]
\centering
\begin{tabular}{ccc}
\hline
Empirical power ($\%$) & & mean (sd) of Hausdorff distance\\
 \cmidrule(lr){1-1} \cmidrule(lr){3-3} 
100 & & 2.19 (1.55) \\ 
\hline
\end{tabular}
\caption{Simulation results of \citet{safikhani2020joint} under the setting in Section \ref{sim_stronger_1}.}
\label{Tab:stronger_setting_ss}
\end{table}


\subsubsection{Low rank + sparse $A^{(1)}$} \label{LS_a1}

\begin{figure}[h!]
\begin{center}
\includegraphics[width=10cm, height=4cm]{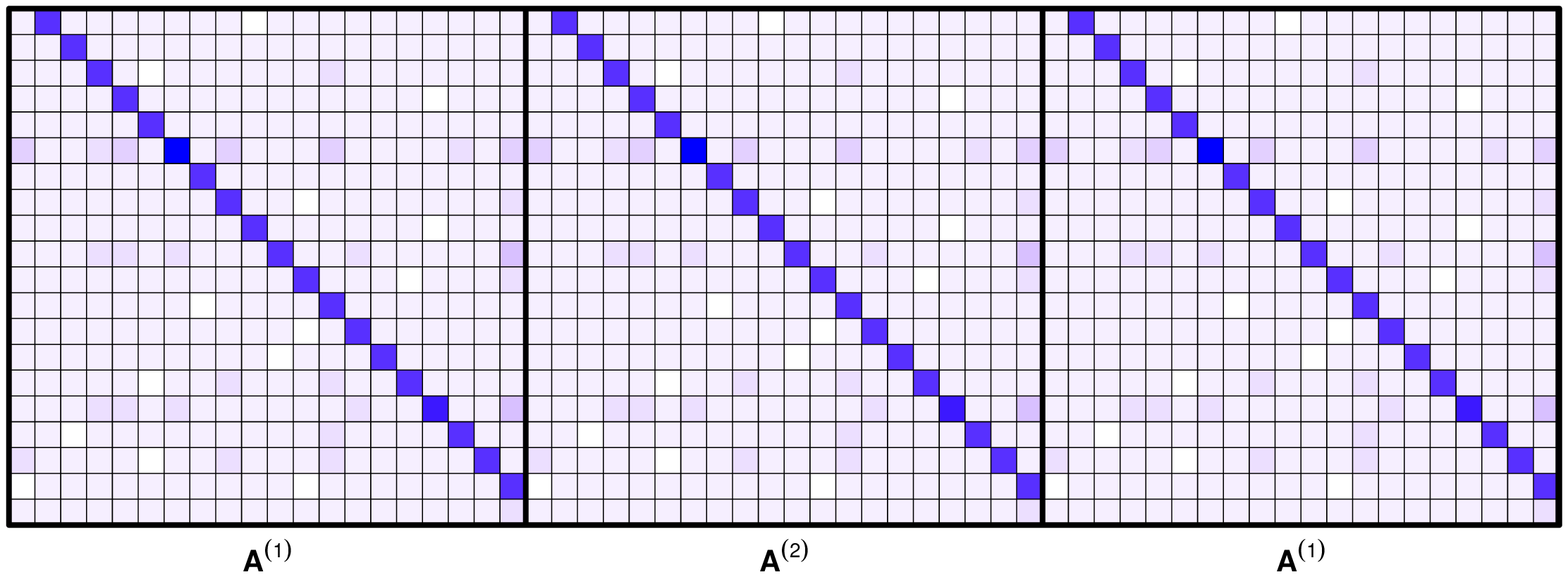}
\end{center}
\caption{The underlying coefficient matrices, $(A^{(1)}, A^{(2)}, A^{(1)})$, for the simulation setting in Section \ref{LS_a1}, where $A^{(2)}$ corresponds to an anomaly.}
\label{fig:single_a_sps}
\end{figure}

\begin{table}[!ht] 
\centering
\begin{tabular}{cccccccc} 
  \hline
T & p &  $[\eta_1, \eta_2]$ & $\eta_2-\eta_1$ & $\text{nzr}(A^{(1)}_\text{sps})$ & $\text{nzr}(A^{(2)}_\text{sps})$  & Range(low-rank) & $\Sigma_{\varepsilon}$ \\
  \hline
300 & 20 & $[T(1/3), T(2/3)]$ & $100$ & $-0.8682672$ & $0.8682672$ & $(-0.217, 0.212)$ & $0.01\textbf{\textit{I}}_p$\\
  \hline
\end{tabular} 
\caption{Simulation setting for Section \ref{LS_a1}, where $\text{nzr}(A^{(1)}_\text{sps})$ and $\text{nzr}(A^{(2)}_\text{sps})$ are the non-zero elements in sparse part of $A^{(1)}$ and $A^{(2)}$, respectively and Range(low-rank) is the range of non-zero elements in low-rank part.}
\label{fig:single_a_sps}
\end{table} 

In this section, we borrow the simulation setting of scenario $A.2$ used in \citet{bai2020multiple} to compare their performance with ours. The underlying VAR coefficient matrix has the low rank plus sparse structure and only the sparse part (i.e. 1-off diagonal in this setting) undergoes change at anomaly. The true coefficient matrices are presented in Figure \ref{fig:single_a_sps}. As all elements of $A^{(1)}$ are non-zero, we estimate $A^{(1)}$ by using the ridge penalty when $A^{(1)}$ is assumed to be unknown.


\begin{table}[h!]
\centering
\begin{tabular}{ccccc}
  \hline
  & & & $A^{(1)}$ is known & $A^{(1)}$ is estimated\\
 \hline
 \multirow{4}{*}{$\Sigma_{\varepsilon}$ is known} & random  & OLS  &  100 & 100  \\   
            &  (s = 845) & Lasso &  100 & 100  \\   
            \cline{2-5}  
            & deterministic  & OLS  & 100 & 100  \\   
            &  (s = 845) & Lasso &  100 & 100  \\    
            \hline
 \multirow{4}{*}{$\Sigma_{\varepsilon}$ is unknown} & random  & OLS  & 100 & 100  \\   
            &  (s = 941) & Lasso &  100 & 100  \\   
            \cline{2-5}  
            & deterministic  & OLS  & 100 & 100  \\   
            &  (s =  941) & Lasso &  100 & 100  \\   
            \hline      
\end{tabular}
\caption{Empirical power ($\%$) from 100 simulation runs for the settings described in Section \ref{LS_a1}, where $s$ is the number of intervals examined.}
\label{Tab:stronger_setting_ours1}
\end{table}

\begin{table}[h!]
\centering
\begin{tabular}{ccccc}
  \hline
  & & & $A^{(1)}$ is known & $A^{(1)}$ is estimated\\
 \hline
  \multirow{4}{*}{$\Sigma_{\varepsilon}$ is known} & random  & OLS  &  0.67 (0.00)  &  1.47 (0.52) \\ 
            &  (s = 845) & Lasso & 0.67 (0.00) & 1.55 (0.66) \\   
            \cline{2-5}  
            & deterministic  & OLS  & 0.33 (0.00) & 0.33 (0.00)\\ 
            &  (s =  845) & Lasso &  0.33 (0.00) & 0.34 (0.07) \\   
            \hline
 \multirow{4}{*}{$\Sigma_{\varepsilon}$ is unknown} & random  & OLS  & 0.67 (0.00) & 1.74 (0.63)  \\   
            &  (s = 941) & Lasso &  0.67 (0.00) & 1.31 (0.55) \\   
            \cline{2-5}  
            & deterministic  & OLS  & 0.61 (0.13) & 0.65 (0.08)  \\   
            &  (s =  941) & Lasso &  0.38 (0.12) & 0.44 (0.15) \\   
            \hline     
  \multicolumn{3}{c}{\citet{bai2020multiple}}  & \multicolumn{2}{c}{2.11 (1.59)} \\ 
  \hline  
\end{tabular}
\caption{The mean (standard deviation) of Hausdorff distance from 100 simulation runs for the settings described in Section \ref{LS_a1}, where $s$ is the number of intervals examined.}
\end{table}

\begin{table}[h!]
\centering
\begin{tabular}{cc}
\hline
\multicolumn{2}{c}{$\#$ of estimated change points from 100 runs} \\
\hline
2 & 3 \\
\hline
80 & 20 \\ 
\hline
\end{tabular}
\caption{Simulation results of \citet{bai2020multiple} under the setting in Section \ref{LS_a1}.}
\label{Tab:stronger_setting_LS}
\end{table}

\pagebreak
\subsection{Simulation results when $\Sigma_{\varepsilon}$ is estimated} \label{sigma_estimated}
In this section, we repeat the simulation settings introduced in Section \ref{sec4} for the case when $\Sigma_{\varepsilon}$ is unknown. 
We use the maximum likelihood estimator for $\hat{\Sigma}_{\varepsilon}$ and compare the performance of our method with the change-point detection technique proposed by \citet{safikhani2020joint}).
As mentioned earlier, the default tuning parameters recommended by \citet{safikhani2020joint} do not fit well in the simulation settings presented in the following sections.
Among three tuning parameters $\lambda_1$, $\lambda_2$ and $\omega$, we adjust $\lambda_1$ and $\omega$; a larger range is examined for finding the optimal $\lambda_1$ in the initial break detection stage and a larger value of  $\omega=4 \log T \log p$ is used (instead of the default constant $1/1.75$) to allow smaller number of break points to be selected in the screening stage.


\subsubsection{Dense $A^{(1)}$, Single anomaly}
\begin{table}[h!]
\centering
\begin{tabular}{ccccc}
  \hline
 & & & $A^{(1)}$ is known & $A^{(1)}$ is estimated\\
 \hline
 \multirow{6}{*}{case 1} & random  & OLS  &  25  &   13 \\ 
            &  (s = 1969) & Lasso &  \textbf{100} & \textbf{78}  \\   
            \cline{2-5}  
            & deterministic  & OLS  & 20 &  29 \\ 
            &  (s = 1969) & Lasso &  \textbf{100} &  \textbf{83}\\   
            \cline{2-5}
            & deterministic  & OLS  &  19  &  28  \\ 
            &  (s = 981) & Lasso &  \textbf{100} &  \textbf{82} \\     
            \hline
\multirow{6}{*}{case 2} & random  & OLS  &  7 & 6 \\ 
            &  (s = 1969) & Lasso &  \textbf{99} &  \textbf{34} \\  
            \cline{2-5}
            & deterministic  & OLS  &  10 &  14\\ 
            &  (s = 1969) & Lasso & \textbf{100} &  \textbf{34}\\        
            \cline{2-5}
            & deterministic  & OLS  &   10 &  13  \\ 
            &  (s = 981) & Lasso &  \textbf{100} &   \textbf{32} \\       
\hline
\end{tabular}
\caption{Empirical power ($\%$) from 100 simulation runs for the settings described in Section \ref{sec.s.a}, where $s$ is the number of intervals examined.}
\label{tab:single_a_count_sigmahat}
\end{table}

\begin{table}[h!]
\centering
\begin{tabular}{rrrrrrrrrrrrrrr}
  \hline
  & \multicolumn{14}{c}{\# of estimated change points}\\
  \cline{2-15}
 & 3 & 4 & 5 & 6 & 7 & 8 & 9 & 10 & 11 & 12 & 13 & 14 & 15 & 17 \\ 
  \hline
case 1 & 0 &  1 &   4 &   6 &  16 &  $\textbf{19}$ &  11 &  10 &  13 &  10 &   4 &   3 &   1 &   2 \\ 
case 2 & 1 &   1 &   7 &  14 &  11 &  $\textbf{25}$ &  10 &  15 &   7 &   4 &   3 &   1 & 0 &  1 \\ 
   \hline
\end{tabular}
\caption{Distribution of the number of estimated change-points by \citet{safikhani2020joint} under the simulation setting described in Section \ref{sec.s.a} over 100 simulation runs.}
\end{table}


\begin{table}[h!]
\centering
\begin{tabular}{ccccc}
  \hline
 & & & $A^{(1)}$ is known & $A^{(1)}$ is estimated\\
 \hline
 \multirow{6}{*}{case 1} & random  & OLS  & 35.41 (18.24)  & 43.00 (8.79)  \\ 
            &  (s = 1969) & Lasso &  2.67 (3.85) & 18.28 (17.75)  \\  
            \cline{2-5}  
            & deterministic  & OLS  & 37.72 (16.39) & 35.73 (17.66) \\  
            &  (s = 1969) & Lasso & 1.84 (4.32) & 12.23 (18.11) \\      
            \cline{2-5}
            & deterministic  & OLS  &  38.33 (15.92)  & 36.62 (16.94) \\ 
            &  (s = 981) & Lasso &  2.00 (4.33) & 13.95 (19.05) \\ 
            \hline
\multirow{6}{*}{case 2} & random  & OLS  & 43.86 (11.50)   & 45.83 (6.16) \\ 
            &  (s = 1969) & Lasso & 3.74 (6.07) & 36.70 (17.51) \\ 
            \cline{2-5}
            & deterministic  & OLS  & 42.58 (13.38) & 42.47 (13.23) \\ 
            &  (s = 1969) & Lasso & 2.50 (5.15) &  36.80 (18.29) \\       
            \cline{2-5}
            & deterministic  & OLS  & 42.80 (13.06) & 42.97 (12.47) \\ 
            &  (s = 981) & Lasso & 2.28 (4.70) & 37.15 (18.01) \\      
\hline
\end{tabular}
\caption{The mean (standard deviation) of Hausdorff distance from 100 simulation runs for the settings described in Section \ref{sec.s.a}, where $s$ is the number of intervals examined.}
\label{tab:single_a_meansd_sigmahat}
\end{table}

\begin{table}[h!]
\centering
\begin{tabular}{cc}
  \hline
 & mean (sd) of Hausdorff distance \\ 
   \hline
case 1 & 22.84  (4.64)\\ 
case 2 & 23.46  (4.45)\\ 
   \hline
\end{tabular}
\caption{The mean (standard deviation) of Hausdorff distance from 100 simulation runs obtained by \citet{safikhani2020joint} under the simulation setting described in Section \ref{sec.s.a}.}
\end{table}

\pagebreak
\subsubsection{Dense $A^{(1)}$, Two anomalies}

\begin{table}[h!]
\centering
\begin{tabular}{rrrrrrrrrrr}
\hline
  & \multicolumn{10}{c}{\# of estimated change points}\\
  \cline{2-11}
 & $\mathbf{4}$ & 5 & 6 & 7 & 8 & 9 & 10 & 11 & 12 & 13 \\ 
  \hline
case 1 &   5 &   5 &  16 &  18 &  $\mathbf{21}$ &  18 &  12 &   4 &   1 & 0 \\ 
case 2 &   4 &  12 &  11 &  21 &  $\mathbf{24}$ &  13 &   7 &   5 &   2 &   1 \\ 
   \hline
\end{tabular}
\caption{Distribution of the number of estimated change points by \citet{safikhani2020joint} under the simulation setting described in Section \ref{ta} over 100 simulation runs.}
\end{table}

\begin{table}[h!]
\centering
\begin{tabular}{cccccccccccc}
  \hline
 & & & \multicolumn{4}{c}{$A^{(1)}$ is known} & & \multicolumn{4}{c}{$A^{(1)}$ is estimated} \\
 \cmidrule(lr){4-7} \cmidrule(lr){9-12} 
 & & & 0 & 1 & \textbf{2} & 3  & & 0 & 1 & \textbf{2} & 3 \\
\hline
\multirow{6}{*}{case 1} & random  & OLS  & \textbf{55} & 45 & 0 & 0 && 48 & \textbf{49} & 3 & 0 \\ 
            &  (s = 1969) & Lasso & 0 & \textbf{69} & 31 & 0 && 5 & \textbf{94}  & 1 & 0 \\ 
            \cline{2-12}  
            & deterministic  & OLS  & 23 & \textbf{65} & 12 & 0 && 37 & \textbf{54} & 9 & 0 \\ 
            &  (s = 1969) & Lasso & 0 & \textbf{65} & 33 & 2 && 5 & \textbf{92} & 3  & 0 \\ 
            \cline{2-12}
            & deterministic  & OLS  & 28 & \textbf{61} & 11  & 0 && 42 & \textbf{55} & 3  & 0 \\ 
            &  (s = 981) & Lasso & 0 & \textbf{66} & 34 & 0 && 5 & 91 & 4 & 0 \\ 
            \hline
 \multirow{6}{*}{case 2} & random  & OLS  & \textbf{72} & 28 & 0 & 0  && \textbf{86} & 14 & 0 & 0 \\ 
            &  (s = 1969) & Lasso & 0 & 5 & \textbf{93} & 2 && 38 & \textbf{62} & 0 & 0 \\ 
            \cline{2-12}  
            & deterministic  & OLS  & \textbf{71} & 23 & 6 & 0 &&  \textbf{75} & 23 & 2 & 0 \\ 
            &  (s = 1969) & Lasso & 0 & 5 & \textbf{95} & 0 && 36 & \textbf{64} & 0 & 0 \\ 
            \cline{2-12}
            & deterministic  & OLS  & \textbf{81} & 16 & 3 & 0 && \textbf{82} & 17  & 1 & 0 \\ 
            &  (s = 981) & Lasso & 0 & 6 & \textbf{94} & 0 && 36 & \textbf{64} & 0 & 0 \\ 
            \hline
\end{tabular}
\caption{Distribution of the number of detected anomalies for two methods in all cases described in Section \ref{ta} over 100 simulation runs, where $s$ is the number of intervals examined.}
\label{tab:two_a_count_sigmahat}
\end{table}

\begin{table}[h!]
\centering
\begin{tabular}{cc}
  \hline
 & mean (sd) of Hausdorff distance \\ 
   \hline
case 1 & 31.42  (3.70) \\ 
case 2 & 20.25  (7.08) \\
   \hline
\end{tabular}
\caption{The mean (standard deviation) of Hausdorff distance from 100 simulation runs obtained by \citet{safikhani2020joint} under the simulation setting described in Section \ref{ta}.}
\end{table}

\begin{table}[h!]
\centering
\begin{tabular}{ccccc}
  \hline
 & & & $A^{(1)}$ is known & $A^{(1)}$ is estimated\\
 \hline
 \multirow{6}{*}{case 1} & random  & OLS  & 29.74 (8.38)  & 30.01 (7.85)   \\ 
            &  (s =  1969) & Lasso &  7.27 (4.38) & 19.02 (9.87)\\  
            \cline{2-5}  
            & deterministic  & OLS  & 28.51 (10.91)  & 29.69 (9.50)\\  
            &  (s =   1969) & Lasso & 5.95 (3.33) & 17.81 (12.08) \\      
            \cline{2-5}
            & deterministic  & OLS  & 29.02 (10.45) & 30.96 (7.39) \\ 
            &  (s =  981) & Lasso & 5.72 (3.16) & 18.05 (12.16) \\ 
            \hline
\multirow{6}{*}{case 2} & random  & OLS  & 13.60 (0.18)   & 13.54 (0.72)\\ 
            &  (s =  1969) & Lasso & 3.42 (5.13) & 12.41 (1.68) \\ 
            \cline{2-5}
            & deterministic  & OLS  & 13.14 (2.77) & 13.44 (1.20) \\ 
            &  (s = 1969) & Lasso & 2.35 (2.71) & 12.72 (1.64) \\       
            \cline{2-5}
            & deterministic  & OLS  & 13.27 (1.76) & 13.41 (1.17) \\ 
            &  (s =  981) & Lasso & 2.47 (2.89) & 12.62 (1.76) \\              
            \hline
\end{tabular}
\caption{The mean (standard deviation) of Hausdorff distance from 100 simulation runs for the settings described in Section \ref{ta}, where $s$ is the number of intervals examined.}
\label{tab:two_a_meansd_sigmahat}
\end{table}

\pagebreak
\subsubsection{Sparse $A^{(1)}$, Single anomaly}


\begin{table}[h!]
\centering
\begin{tabular}{ccccc}
  \hline
 & & & \multicolumn{2}{c}{$\#$ anomaly detection from 100 runs}  \\
 \cline{4-5}
 & & & $A^{(1)}$ is known & $A^{(1)}$ is estimated\\
 \hline
 \multirow{4}{*}{case 1} & random  & OLS  &  2 &   3 \\ 
            &  (s = 469) & Lasso &  \textbf{100} & \textbf{99}  \\   
            \cline{2-5}  
            & deterministic  & OLS  & 73 &  73 \\ 
            &  (s =  469) & Lasso &  \textbf{100} &  \textbf{100}\\   
            \hline
 \multirow{4}{*}{case 2} & random  & OLS  &  2 &  1 \\ 
            &  (s = 469) & Lasso &  \textbf{85} & \textbf{33}  \\   
            \cline{2-5}  
            & deterministic  & OLS  & 57 &  \textbf{66} \\ 
            &  (s =  469) & Lasso &  \textbf{100} &  52 \\   
            \hline
\end{tabular}
\caption{Empirical power ($\%$) from 100 simulation runs for the settings described in Section \ref{sec4.2.1}, where $s$ is the number of intervals examined.}
\end{table}

\begin{table}[h!]
\centering
\begin{tabular}{rrrrrrr}
  \hline
  & \multicolumn{6}{c}{\# of estimated change points}\\
  \cline{2-7}
 & 0 & 1 & 2 & 3 & 4 & 5 \\ 
  \hline
case 1  &   1 &  34 &  \textbf{51} &  12 &   1 &   1 \\ 
case 2  &  1 &  29 &  \textbf{56} &  12 &   1 &   1 \\ 
   \hline
\end{tabular}
\caption{Distribution of the number of estimated change points by \citet{safikhani2020joint} under the simulation setting described in Section \ref{sec4.2.1} over 100 simulation runs.}
\end{table}

\begin{table}[h!]
\centering
\begin{tabular}{ccccc}
  \hline
 & & & \multicolumn{2}{c}{mean (sd) of Hausdorff distance}  \\
 \cline{4-5}
 & & & $A^{(1)}$ is known & $A^{(1)}$ is estimated\\
 \hline
 \multirow{4}{*}{case 1} & random  & OLS  & 44.44 (2.55) & 43.92 (6.08) \\ 
            &  (s = 469) & Lasso & 1.50 (0.67) & 1.75 (4.36) \\   
            \cline{2-5}  
            & deterministic  & OLS  & 20.72 (19.56) & 23.43 (19.20)\\ 
            &  (s = 469) & Lasso & 0.32 (0.15) & 0.32 (0.15)\\   
            \hline
 \multirow{4}{*}{case 2} & random  & OLS  & 45.93 (3.47) & 46.35 (0.50)   \\ 
            &  (s = 469) & Lasso & 12.97 (14.19) & 31.43 (21.44) \\   
            \cline{2-5}  
            & deterministic  & OLS  & 29.41 (19.33) & 30.27 (17.77) \\ 
            &  (s = 469) & Lasso & 0.46 (0.15) & 22.50 (23.08) \\   
            \hline
\end{tabular}
\caption{The mean (standard deviation) of Hausdorff distance from 100 simulation runs for the settings described in Section \ref{sec4.2.1}, where $s$ is the number of intervals examined.}
\end{table}

\begin{table}[h!]
\centering
\begin{tabular}{cc}
  \hline
 & mean (sd) of Hausdorff distance \\ 
   \hline
case 1 & 45.65  (1.84)\\ 
case 2 & 45.82  (2.46)\\ 
   \hline
\end{tabular}
\caption{The mean (standard deviation) of Hausdorff distance from 100 simulation runs obtained by \citet{safikhani2020joint} under the simulation setting described in Section \ref{sec4.2.1}.}
\end{table}

\end{document}